\newif\ifShowKeys
\ShowKeystrue

\documentclass[11pt]{article}
	\pdfoutput=1
\usepackage[T1]{fontenc}

\usepackage{listings}
\lstnewenvironment{arkady}  
{\lstset{language=C,frame=trbl,basicstyle = \footnotesize \ttfamily , breaklines = true,showstringspaces=false}}{}

\usepackage[usenames,dvipsnames]{xcolor}
\usepackage[setpagesize=false,pagebackref=false, 
linktocpage, bookmarksopen=true, colorlinks=true, 
linkcolor=Maroon,citecolor=Maroon,urlcolor=Maroon]{hyperref}

\usepackage[parsep]{collref}	

\usepackage{amsmath, amssymb,amsthm}
\usepackage{stackrel}
\numberwithin{equation}{section}
\usepackage{bm,environ,mathrsfs,array,arydshln}
\usepackage{booktabs,float,slashed}
\usepackage{appendix}
\usepackage[mathcal]{euscript}
\usepackage{tensor} 						
\usepackage{mathabx}
\usepackage[vcentermath]{youngtab}

\usepackage{graphicx,epsfig,epic}
\usepackage{tikz} 
\usetikzlibrary{arrows,decorations.pathreplacing,decorations.markings,snakes}
\usetikzlibrary{cd}

\usepackage{datetime}
\usepackage{comment}					

\allowdisplaybreaks

\usepackage{framed}						
\definecolor{shadecolor}{rgb}{0.9996078, 0.984314, 0.960784}
\definecolor{framecolor}{rgb}{0,0,0}
\definecolor{TFTitleColor}{RGB}{1,1,1}
\definecolor{TFFrameColor}{RGB}{249	218	181}		
\definecolor{TFFrameColor}{RGB}{230 230 230 }

\newenvironment{frshaded}{%
    \MakeFramed {\FrameRestore}}%
    {\endMakeFramed}

\definecolor{myred}{RGB}{233, 33, 45}

\newcommand{\bs}{\begin{frshaded}}			
\newcommand{\es}{\end{frshaded}\noindent}

\def\ba#1\ea{\begin{align}#1\end{align}}		        
\newcommand{\be}{\begin{equation}}
\newcommand{\ee}{\end{equation}}
\newcommand{\bea}{\begin{equation} \begin{aligned}} 
\newcommand{\eea}{\end{aligned} \end{equation}}
\newcommand{\mc}{\mathcal }

\newcommand{\la}{\label}
\newcommand{\eps}{\varepsilon}

\newcommand{\lp}{\notag \\ & }

\DeclareMathOperator{\vol}{vol}

\newcommand{\cf}{\textit{cf.} }
\newcommand{\ie}{\textit{i.e.} }
\newcommand{\eg}{\textit{e.g.} }

\newcommand{\N}{\mathcal N}

\newcommand{\sql}{\sqrt\l}
\renewcommand{\l}{\lambda}

\newcommand{\sym}{\text{\tiny SYM}}
\newcommand{\abjm}{\text{\tiny ABJM}}
\newcommand{\vev}[1]{\langle #1 \rangle}
\newcommand{\ads}{\text{AdS}}
\newcommand{\Z}{\mathbb{Z}}
\newcommand{\vp}{\varphi}
\newcommand{\T}{\text{T}}
\newcommand{\KK}{\mathbb{K}}
\newcommand{\EE}{\mathbb{E}}
\newcommand{\cp}{\mathbb{CP}^{3}}
\newcommand{\wb}{\bar w}
\newcommand{\kk}{\eps}

\DeclareMathOperator{\cn}{cn}
\DeclareMathOperator{\sn}{sn}
\DeclareMathOperator{\dn}{dn}

\newcommand{\foot}{\footnote}
\newcommand{\ci}{\cite}
\def\ov{\over}
\newcommand{\rf}[1]{(\ref{#1})}
\def\no{\nonumber}

\def \vol {{\rm vol}}
\def\a{{\alpha}}
\def \ha {{1 \ov 2}}
\def \te {\textstyle}
\def \l {\lambda} 
\def \no {\notag}
\def \iffa {\iffalse} 
\def \ed {\small
\bibliography{BT-Biblio}
\bibliographystyle{JHEP-v2.9}
\end{document}}

\def \la {\label}
 
\def \l {\lambda} 
\def \gs  {g_{\rm s}}  \def \CP  {{\rm CP}}
\def \cp  {{\rm CP^3}}

\def \ha {{1\ov 2}}

\def \te {\textstyle}

\begin{document}



\begin{titlepage}

\vspace*{15mm}
\begin{center}
{\Large\sc   Non-planar corrections to ABJM Bremsstrahlung function \\
\vskip 8pt
from quantum M2 brane}

\vspace*{10mm}

{\Large M. Beccaria${}^a$, A. A. Tseytlin${}^b$\footnote{\ Also at the Institute for Theoretical and Mathematical Physics (ITMP) of MSU  and Lebedev Institute.}} 

\vspace*{4mm}
	
${}^a$ Universit\`a del Salento, Dipartimento di Matematica e Fisica \textit{Ennio De Giorgi},\\ 
		and INFN - sezione di Lecce, Via Arnesano, I-73100 Lecce, Italy
			\vskip 0.3cm
${}^b$ Blackett Laboratory, Imperial College London SW7 2AZ, U.K.
			\vskip 0.3cm
\vskip 0.2cm {\small E-mail: \texttt{matteo.beccaria@le.infn.it}, \texttt{tseytlin@imperial.ac.uk}}
\vspace*{0.8cm}
\end{center}

\begin{abstract}  
As was shown in arXiv:2303.15207, the leading large $N$, fixed $k$ correction 
in the localization result for the expectation value of the $\ha$ BPS  circular 
Wilson loop in  $U(N)_{k}\times U(N)_{-k}$ ABJM theory  given   by the $(\sin{2\pi\ov k})^{-1}$  factor
can be reproduced on the dual M-theory side  as the one-loop  correction  in the partition function of     M2   brane  in AdS$_{4}\times S^{7}/\mathbb{Z}_{k}$  with AdS$_{2}\times S^{1}$ world volume. 
Here we prove, following the suggestion in  arXiv:2408.10070, that the analogous 
fact  is true also for  the  corresponding  correction $B_1=-\frac{1}{2\pi k}\cot\frac{2\pi}{k}$ in the localization result for the Bremsstrahlung function
associated with the Wilson loop  with a small cusp in either AdS$_4$ or  $\CP^3$. The corresponding  M2  brane   is wrapped on the 11d circle  and  generalizes the type IIA 
string solution in AdS$_{4}\times \CP^3$  ending on the cusped line. 
We show  that the one-loop term in the 
M2 brane partition function reproduces the localization
expression for $B_1$ as the coefficient of 
 the leading term  in its   small cusp   expansion. 
\end{abstract}
\vskip 0.5cm
	{
	}
\end{titlepage}

\iffa

As was shown in arXiv:2303.15207,  the leading large $N$, fixed $k$ correction 
in the localization result for the expectation value of the $\frac{1}{2}$ BPS  circular 
Wilson loop in  $U(N)_{k}\times U(N)_{-k}$ ABJM theory  given   by the $(\sin\frac{2\pi}{ k})^{-1}$ 
 factor can be reproduced on the dual M-theory side  as the one-loop  correction  in the partition function of     an M2   brane  in AdS$_{4}\times S^{7}/\mathbb{Z}_{k}$  with AdS$_{2}\times S^{1}$ world volume.  Here we prove, following the suggestion in  arXiv:2408.10070, that the analogous 
fact  is true also for  the  corresponding  correction $B_1=-\frac{1}{2\pi k}\cot\frac{2\pi}{k}$ in the localization result for the Bremsstrahlung function associated with the Wilson line  with a small cusp in either AdS$_4$ or  $\rm CP^3$. The corresponding  M2  brane   is wrapped on the 11d circle  and  generalizes the type IIA  string solution in AdS$_{4}\times \rm CP^3$  ending on the cusped line. 
We show  that the one-loop term in the M2 brane partition function reproduces the localization
expression for $B_1$ as the coefficient of  the leading term  in its   small cusp   expansion. 

\fi

\tableofcontents
\vspace{1cm}

\def \adsc {AdS$_4\times \CP^3$} \def \cs { } 
\def \adsz  {AdS$_{4}\times S^{7}/\mathbb{Z}_{k}$ }
\def \adsz {AdS$_4 \times S^7/\mathbb Z_k$ }
\def \brem {Bremsstrahlung }
\def \om {\omega}
\def \vp  {\varphi} 
\def \ssigma {{\hat \sigma}}
\def \E {{\cal E}}
\def \del {\partial} \def \s  {\sigma}
\def \TT {{\cal T}}
 \def \s  {\sigma}
\def \cpp { {_{\rm CP}}}
\def \aads {{_{\ads}}}
\def \aa {\alpha}
\def \bb {\beta}

\section{Introduction}

A series of recent papers have demonstrated how non-planar  corrections  in  the ABJM theory   \ci{Aharony:2008ug}  can be found by semiclassically quantizing the M2 brane  
in \adsz\  \cite{Giombi:2023vzu,Beccaria:2023ujc,Giombi:2024itd}
(see  also 
\cite{Drukker:2020swu,Beccaria:2023sph,Drukker:2023jxp,Drukker:2023bip}).
Here  we shall follow \cite{Giombi:2023vzu,Giombi:2024itd} and focus on the 
 \brem function $B(\l,N)$  associated with the 
  $\frac{1}{2}$-BPS Wilson loop  with the aim   to demonstrate how 
the result  found  from localization  can be reproduced on the dual M-theory side by quantizing the M2 brane near the classical solution representing the Wilson line  with   a small  cusp.

\

Let us start with a  brief review of some relevant facts  about 
the \brem function  $B$. It  determines the energy emitted by a moving quark  given by 
$\Delta E = 2\pi B\int dt\, \dot v^{2}$ in the small velocity limit.
In the $\N= 4$ $SU(N)$ SYM theory it  may be found from the exact localization result 
for the expectation value of the $\frac{1}{2}$-BPS circular Wilson loop as \cite{Correa:2012at}
\ba
\la{1.1}
B_{\sym} = \frac{1}{2\pi^{2}}\l\frac{\partial}{\partial\l}\log\vev{W}_{\sym}\ ,
\ea
where \cite{Drukker:2000rr,Pestun:2007rz} 
\be
\vev{W}_{\sym} = e^{\frac{\l}{8N^{2}}(N-1)}L_{N-1}^{(1)}\Big(-\frac{\l}{4N}\Big) = \frac{2N}{\sql}I_{1}(\sql)\Big[1+\frac{1}{N^{2}}
\Big(
\frac{\l^{3/2}}{96}\frac{I_{2}(\sql)}{I_{1}(\sql)}-\frac{\l}{8}\Big)+\cdots\Big]\ .
\ee
Expanded in large $N$ and then also in large  $\l$ this gives
\be
B_{\sym}(\l, N) = B_{\sym}^{
}(\l)+\frac{1}{128\pi^{2}}\frac{\l^{3/2}}{N^{2}}+\cdots, \qquad 
B_{\sym}^{
}(\l) = \frac{\sql}{4\pi^{2}}\frac{I_{2}(\sql)}{I_{1}(\sql)} = \frac{\sql}{4\pi^{2}}-\frac{3}{8\pi^{2}}+\cdots.
\ee
 One may also  get $B$ from  the expression for the $\frac{1}{2}$-BPS Wilson loop wrapped w times on the circle  as 
\cite{Lewkowycz:2013laa}
\be
\la{1.4}
B(\l, N) = \frac{1}{4\pi^{2}}\frac{\partial}{\partial\text{w}}\log\vev{W}\Big|_{\text{w}=1}\ .
\ee
In the $\N= 4$ SYM case this leads to the same expression as in (\ref{1.1}) since the dependence on w can be incorporated into $\langle W\rangle$
 by $\sql\to \text{w}\sql$ \cite{Drukker:2000rr}. 
  Ref.  \cite{Lewkowycz:2013laa} has shown that (\ref{1.4}) applies also
   in the ABJM case for the \brem function given in terms of
 the $\frac{1}{6}$-BPS Wilson loop.
 
The \brem function may also be related to the  anomalous dimension $\Gamma_{\rm cusp}$
governing the logarithmic divergence of a Wilson loop with a cusp $\langle W\rangle \sim \exp[ {-\Gamma_{\rm cusp}\log(\Lambda_{\rm IR}/\Lambda_{\rm UV})}]$.
In the case of locally supersymmetric Wilson loops, the cusp anomaly $\Gamma_{\rm cusp}=\Gamma_{\rm cusp}(\l,N; \aa, \bb)$ 
depends 
on the geometrical angle $\aa$ between the two 
 lines defining the cusp, 
 and an internal angle $\bb$ describing the change in the ``internal''  orientation 
 described by  the scalar  coupling. 
 In a small angle expansion around the BPS limit $\aa = \pm \bb$ 
one can determine  $B(\l,N)$  
 from  the small $\aa, \bb$ expansion of $\Gamma_{\rm cusp}$ as  \cite{Drukker:2011za,Griguolo:2012iq}
\be \la{1.5}
 \Gamma_{\rm cusp}(\l,N;  \aa, \bb) = - (\aa^{2}-\bb^{2})\, B(k, N)+\cdots.
\ee
One may also compute the \brem function corresponding to either $\frac{1}{2}$-
 or $\frac{1}{6}$-BPS Wilson loops in the ABJM theory  by using a generalization of the identity \cite{Correa:2012at} that expresses $B$ as a derivative
 of the logarithm of the latitude Wilson loop with respect to the small latitude angle.\foot{For $\frac{1}{2}$-BPS Wilson loop this identity was proposed and proved perturbatively in \cite{Bianchi:2014laa}, and for the corresponding \brem function 
 it was first introduced and then proved exactly in \cite{Bianchi:2017ozk}.  
 In the $\frac{1}{6}$-BPS Wilson loop case a similar identity 
 for the Brehmstrahlung function was proved in \cite{Correa:2014aga} and further elaborated on in \cite{Bianchi:2018scb}.  
 For a review of the Brehmstrahlung function in the ABJM theory see the contribution of L. Bianchi in \cite{Drukker:2019bev} and also \cite{Penati:2021tfj}.}
 In the planar limit of $U(N)_{k}\times U(N)_{-k}$ ABJM theory  one finds the following strong coupling expansion for the \brem function corresponding to the $\frac{1}{2}$-BPS Wilson 
 loop \cite{Bianchi:2018scb}
 \be
 \la{1.6}
 B^{
 }_{\abjm}(\l)\Big|_{\l \gg 1}  = \frac{1}{2\pi}\sqrt\frac{\l}{2}-\frac{1}{4\pi^{2}}-\frac{1}{96\pi}\frac{1}{\sqrt{2\l}}+\cdots, \qquad
 \l = \frac{N}{k} = \text{fixed}, \ N\to\infty  \ .
 \ee
 This  matches the prediction from string theory in \adsc\ at the two leading orders
 \cite{Forini:2012bb,Aguilera-Damia:2014bqa}.\foot{This was partly shown in  \cite{Forini:2012bb} by computing the one-loop
 contribution to $ \Gamma_{\rm cusp}$  at $\bb=0$ and small $\aa$. They also computed 
 $ \Gamma_{\rm cusp}$
  at $\aa=0$ and small $\bb$ with a result not consistent with the expected  expression in  (\ref{1.5}). This was
 later corrected in  \cite{Aguilera-Damia:2014bqa}.
 }
  Finding non-planar corrections in
 this  approach  is  hard  as  the  exact  localization  result 
  is  not  known in the ABJM theory  for  a  non-zero  cusp  angle. 
 
An  alternative approach based on mass-deformed localization 
matrix model  was suggested in \cite{Guerrini:2023rdw,Armanini:2024kww}.
The resulting  exact expression for \brem function found in \cite{Armanini:2024kww}  reads (for  $k>2$)
 \be \la{1.7}
 B_{\abjm}(k,N) = -\frac{1}{(4 \pi^2k)^{2/3}}\, 
 \frac{\text{Ai}'[(\frac{\pi^{2}}{2}k)^{1/3}(N-\frac{k}{24}-\frac{1}{3k})]}
 {\text{Ai}[(\frac{\pi^{2}}{2}k)^{1/3}(N-\frac{k}{24}-\frac{1}{3k})]}-\frac{1}{2\pi k}\cot\frac{2\pi}{k}\ ,
 \ee
 where prime indicates the derivative of the Airy function over its argument. 
 As was   pointed out in \cite{Giombi:2024itd} the expression (\ref{1.7})
 can be reproduced in a simple way 
  by  applying  (\ref{1.4})  to 
the result for  the localization result for the 
 expectation value of the $\frac{1}{2}$-BPS  circular Wilson  loop 
 in the w-fundamental representation (presumably equivalent to the result for the 
 w-wrapped   circular loop)  \cite{Klemm:2012ii}
\be\la{1.8}
\vev{W}_{\abjm} = \frac{1}{2\sin\frac{2\pi \text{w}}{k}}\frac{\text{Ai}[(\frac{\pi^{2}}{2}k)^{1/3}(N-\frac{k}{24}-\frac{1}{3k}-\frac{2\text{w}}{k})]}
{\text{Ai}[(\frac{\pi^{2}}{2}k)^{1/3}(N-\frac{k}{24}-\frac{1}{3k})]}\ .
\ee
 Expanding  in 
large $N$ at fixed $k$  one gets 
\ba
\la{1.9}
&\vev{W}_{\abjm} = \frac{1}{2\sin\frac{2\pi\text{w}}{k}}e^{\pi\text{w}\sqrt\frac{2N}{k}}\Big[1-\frac{\pi\text{w}(k^{2}+24\text{w}+8)}
{24\sqrt{2}\,k^{3/2}}\frac{1}{\sqrt N}+\cdots\Big]\ ,
\\
\la{1.10}
&B_{\abjm}(k,N) = \frac{1}{4\pi^{2}}\frac{\partial}{\partial\text{w}}\log\vev{W}_\abjm\Big|_{\text{w}=1}= \frac{1}{4\pi}\sqrt\frac{2N}{k}-\frac{1}{2\pi k}\cot\frac{2\pi}{k}-\frac{56+k^{2}}{96\pi\sqrt 2\,k^{3/2}}\frac{1}{\sqrt N}+\cdots\ . \qquad
\ea
The first   term in \rf{1.10}  is same as  in the planar limit in \rf{1.6}.
The cot term in \rf{1.7},\rf{1.10}  originates from the  derivative of the  logarithm of the 
1/sin  prefactor in \rf{1.8}. 
 Expanded in large $k$  it  gives  an 
   infinite series of terms  $1/k^{2p}=(\l/N)^{2p}$ that 
 represent  the leading large $\l$ corrections at each order in $1/N$,  with the 
  first $p=0$ one   reproducing  the  $-{1\ov 4 \pi^2}$ correction in \rf{1.6}.
 

\iffa 
\footnote{
In (\ref{1.9}), the dependence on $\text{w}$  of the tree-level $e^{\pi\text{w}\sqrt\frac{2N}{k}}$ and one-loop $\frac{1}{2\sin\frac{2\pi\text{w}}{k}}$
is actually the same as in the case when the M2 brane is wrapped w times not on the $\ads_{4}$ boundary circle but on the 11d circle $\phi$.  In this case we have 
effectively $\phi\to \text{w}\,\phi$ and thus the radius $1/k$ is rescaled to $\text{w}/k$. This leads to $2\pi/k\to 2\pi\text{w}/k$
in the M2 brane one-loop correction. The w-dependence of the subleading terms in (\ref{1.9}) (two and higher loop M2 brane corrections) 
does not appear to have a similar simple explanation.} 
\fi

\

Let us now  recall  some basic  M-theory  relations used in  \cite{Giombi:2023vzu}.
The M2 brane is placed into  $\ads_{4}\times S^{7}/\Z_{k}$  background 
with the metric 
\ba
\la{1.11}
ds^{2} &= \tfrac{1}{4}R^{2}ds^{2}_{\ads_{4}}+R^{2}ds^{2}_{S^{7}/\Z_{k}}\ ,  \\
\la{1.12}
ds^{2}_{\ads_{4}} &= -\cosh^{2}\rho\, dt^{2}+d\rho^{2}+\sinh^{2}\rho\, (dx^{2}+\cos^{2}x\, d\theta^{2})\ , \\
\la{1.13}
ds^{2}_{S^{7}/\Z_{k}} &= ds^{2}_{\cp}+\frac{1}{k^{2}}(d\phi+k A)^{2},\qquad  \qquad \phi\equiv \phi+2\pi\ , \\
\la{1.14}
ds^{2}_{\cp} &= \frac{dw^{s}d\wb^{s}}{1+|w|^{2}}-\frac{w_{r}\wb_{s}dw^{s}d\wb^{r}}{(1+|w|^{2})^{2}}\ , \qquad
A = \frac{i}{2}\frac{w^{s}d\wb^{s}-\wb^{s}dw^{s}}{1+|w|^{2}}\ , \ \ \ \ s,r=1,2,3\ ,
\ea
and the  4-form field strength  
\be\la{1.15}
F_{4} = dC_{3} = -\tfrac{3}{8}R^{3}\vol_{\ads_{4}}\ .
\ee
The leading order relation
 between the radius (in 11d Planck length units) and the parameters
$N,k$ of the dual ABJM theory is\footnote{\la{foot:hirano}As in \cite{Giombi:2023vzu,Giombi:2024itd}  
  we shall  ignore the  shift   \cite{Bergman:2009zh}   $N \to N - {1\over 24} (k - k^{-1})$ as  it will not 
 be relevant to the order of the large $N$  expansion  that  we will consider.}
\be\la{1.16}
\Big(\frac{R}{\ell_{P}}\Big)^{6} = 2^{5}\pi^{2}\, Nk\ .
\ee
The world-volume action for a probe M2 brane in this background is \cite{Bergshoeff:1987cm,deWit:1998yu}\foot{We  choose   the sign  of the action as appropriate for a Euclidean  continuation that will be implicitly assumed below.}
\be\la{1.17}
S_{\rm M2} = T_{2}\int d^{3}\sigma\sqrt{-g}+T_{2}\int C_{3}+\text{fermionic terms},\qquad \qquad 
T_{2} = \frac{1}{(2\pi)^{2}\ell_{P}^{3}} \ . 
\ee
The resulting dimensionless   effective  M2 brane tension  is
\be\la{1.18}
  \T_{2} \equiv  R^{3}T_{2} = \tfrac{1}{\pi}\sqrt{2k N}\ .
\ee
This  suggests  that the expansions in \rf{1.9} and \rf{1.10}  should be matched with   the  semiclassical  (large $\T_2$ at fixed $k$)  expansions  of  the corresponding M2  brane 
expressions. 

Note that  expressed in terms of the  type IIA  string  effective tension $T$   and  coupling $\gs$ 
\be T ={R^2\ov 8 \pi \a'} = \sqrt{\l\ov2} \ , \qquad \ \ 
  \gs  = {   \sqrt \pi\,(2 \l)^{5/4}\ov  N}  \ , \qquad   \ \ 
     \l = {N \ov k}  \ , \ \ \ \ \qquad   {1\ov k^2} = { \gs^2 \ov 8 \pi T}\ ,  \la{1.199}
\ee
the subleading  cot term  in \rf{1.10} represents the sum of the leading 
large  tension corrections at each order in the  string coupling (genus)   expansion 
\be 
-\frac{1}{2\pi k}\cot\frac{2\pi}{k}
= -\frac{1}{4 \pi ^2}+\frac{\gs^2}{24 \pi\,  T}+\frac{\gs^4}{720\, T^2}+\frac{\pi  \gs^6}{15120\, T^3}+\cdots \ . \la{1.200}\ee
Subleading in large $T$ terms at each order in $\gs^2$  should  come from the next $1/\sqrt N$ term in \rf{1.10} or  the  2-loop M2  brane  correction.

As was   shown in \cite{Giombi:2023vzu}  for w=1 
the  exponent and the 1/sin prefactor in \rf{1.9} can be reproduced  on the dual 
M-theory side  as, respectively,  the classical and the  one-loop  corrections 
 in the   partition function  for the M2  brane  in \adsz  wrapped  on  11d circle and 
 AdS$_2 \subset$ AdS$_4$ (ending on a circle at the boundary). 
 
 The  aim of the present  paper will be to show that one can  similarly 
reproduce the  first two terms in  $B$ in (\ref{1.10}) 
from   the  classical   and one-loop  corrections to the M2  brane partition 
 function computing  the cusp anomaly \rf{1.5}. 
 The  corresponding M2 brane solution   will be a  straightforward generalization 
 of  the type IIA string solution    for the line with a  cusp  
  \ci{Drukker:2007qr,Aguilera-Damia:2014bqa}
  wrapped also on the 11d circle.

\

The  straight-line  Wilson loop
  is described, like in  \cite{Giombi:2023vzu},   by the $\ads_{2}\times S^{1}$  
 M2  brane  solution. In Poincare coordinates for the AdS$_4$  the 
 $\ads_{2}$  metric is $z^{-2} ( -dt^2 + dz^2$) with $t$ parametrizing the line.  
Here we will  consider 
the M2 brane  solution 
representing  two lines with a relative angle $\alpha$  in AdS$_4$ 
and  angle $\beta$ in  $\CP^3$  and wrapped also on the  angle $\phi$ in  \rf{1.13}. 
It is  given by the  straightforward uplift of  the IIA  string  solution
 in $\ads_{4}\times \CP^3$
ending on a cusped line \cite{Drukker:2007qr}. The IIA 
 string world sheet  is embedded into 
 a  subspace $\ads_{3}\times S^{1}$ with 
an angle of  $\ads_{3}$ spanning the range $[\frac{\alpha}{2},\pi-\frac{\alpha}{2}]$ corresponding to the directions of the two 
half-lines  representing the  cusp with a non-zero  angle $\alpha$.
 The coordinate of $S^{1}\subset \cp$   belongs to  the interval   $[-\frac{\beta}{2},\frac{\beta}{2}]$  where $\beta$ is   
the  ``internal''  cusp angle.

The corresponding classical  action of the M2 brane  will be proportional to ${1\ov k} \T_{2}$ 
and  will match  the first term in (\ref{1.10}). 
 Quantum M2 brane  fluctuations around this  classical 
solution will   reproduce the $\mc O(\T_{2}^{0})$  term  cot${2\pi \over k}$   in (\ref{1.10}). The computation of this one-loop M2  brane correction  will be the aim of this paper.

In general, the M2 brane partition function will have the form 
\ba
\la{1.19}
Z = \int[dX d\vartheta]\ e^{-S[X,\vartheta]} = & \mc Z_{1}\, e^{-\T_{2}\, \bar{S}_{\rm cl}}\big[1+\mc O(\T_{2}^{-1})\big] = e^{-{\cal T}\, \Gamma_{\rm cusp}} \ ,  
 \\
\la{1.20} 
\mc Z_{1} =& e^{-{\cal T}\, \Gamma^{(1)}_{\rm cusp}}\ ,
\ea
where $(X,\vartheta)$ are the bosonic and fermionic coordinates.
 $\cal T\to \infty $ is the  range of the time direction $t$ parametrizing the cusped line 
 (it plays   the role of an infrared cut off, cf. the discussion above \rf{1.5}). 
The one-loop $\Gamma^{(1)}_{\rm cusp} $    term  should  match the $N^0$ term  in \rf{1.10}  while the   2-loop $\T_{2}^{-1}$   term
 should  reproduce the subleading $N^{-1/2}$ term in \rf{1.10}. 

The one-loop correction $\mc Z_{1}$  is given by the usual combination of  determinants of the 2nd order fluctuation operators. Using  static gauge   and expanding the   M2 brane  3d fluctuation fields in Fourier modes in $S^1$ 
we get  as in   \cite{Giombi:2023vzu}  a tower of 2d   massive fluctuation fields 
with the lowest $n=0$   level  corresponding to the IIA string  fluctuations.

A complication  compared to 
the  circular or straight  Wilson loop case   in   \cite{Giombi:2023vzu}
is that here the induced metric is not just of  a homogeneous AdS$_2 \times S^1$  
space  as in the absence of the cusp 
  and thus the computation of the fluctuation determinants  is, in general,  non-trivial. 
Because of the translation invariance in $t$  the result for $\Gamma^{(1)}_{\rm cusp} $
 can be represented in terms of the vacuum energy of the quadratic fluctuations
 around the classical solution
(here $I$ stands for   the mode number labels)
\be
\la{1.21}
\Gamma^{(1)}_{\rm cusp} \equiv E= 
\tfrac{1}{2}\sum_{ I}(-1)^{\rm F_I}\, \omega_{ I}\ .
\ee
The  fluctuation   energies $\omega_{ I}(\aa, \bb)$ may be evaluated 
in perturbation theory near the BPS limit, i.e. expanding   in the small  cusp angles  like in 
 \cite{Aguilera-Damia:2014bqa}.
 The 
 ``unperturbed'' configuration corresponds to the  straight BPS Wilson line in $\ads_{4}$ (and point-like in $\CP^3$)
 for which the M2  geometry is  AdS$_2\times S^1$
 (in this  case  $E=0$ due to  supersymmetry
which is explicit in the spectrum of fluctuations as in the string case in \cite{Drukker:2000ep,Aguilera-Damia:2014bqa}). \foot{The resulting  procedure of  computing the  quadratic  in small angle  terms in  the one-loop determinants is 
closely related to the  alternative interpretation of the \brem function as
a coefficient in  the 2-point function of the excitations on the Wilson line defect
represented  by  string  or  M2  brane  fluctuations in transverse   directions.
It is also 
  somewhat similar to the one in the 
 near short-string expansions   discussed in \cite{Tirziu:2008fk,Beccaria:2012xm}.}

 As a result, as we  will show  below,   $\Gamma^{(1)}_{\rm cusp}$ takes the form of   \rf{1.5}   with    the one-loop  correction to the \brem function  reproducing the subleading term in \rf{1.10}
   \ba
    B^{(1)} = &  
     -  \frac{1}{ 4\pi^2} + { {2\ov \pi^2}} \sum_{n=1}^\infty {1\ov k^2 n^2 -2}  =- {1\ov 2 \pi k } \cot{ 2 \pi\ov k}\ , \qquad \qquad k>2\ , \la{1.22}\\
B^{(1)} = &  \frac{1}{ 4\pi^2} \ , \ \ \ \ \ \ \ \ \ \  k=1,2 \ . \la{1.23}
\ea
The  $k=1,2$   values  represent the predictions as, like for the BPS Wilson loop,  the corresponding localization results for the \brem function are not currently available. 

Let us note that the result for $ B^{(1)} $ in \rf{1.22},\rf{1.23} is  manifestly finite. 
 This is to be compared with
 the  circular Wilson loop  computation  \cite{Giombi:2023vzu}
 of the one-loop $(\sin {2\pi\ov k})^{-1}$ prefactor  in the   AdS$_2\times S^1$   M2  brane  partition function  where  the sum over $S^1$  mode number $n$ 
 was linearly divergent and thus   required 
the  $\zeta$-function regularization.

\iffa 
Let us recall that  an alternative interpretation of the \brem function is 
that of the coefficient of the 2-point function of the excitations on the Wilson line defect
represented  by  string or  M2  brane  fluctuations in transverse   directions.
Thus one possible generalization of the computation discussed below  is  to 
higher point defect correlation functions as discussed in \cite{Giombi:2017cqn}.
\fi

\

The plan of the rest of this paper is as follows.
In section \ref{sec:classical} we briefly review the classical solution for the  IIA string
 in $\ads_{4}\times\cp$ with a world  sheet ending on a cusped Wilson line with the geometrical angle $\alpha$ in AdS$_4$ 
and the internal angle $\beta$ in $\CP^3$.
 We then   present its 11d uplift as an  M2 brane embedded  in \adsz.
  
In section \ref{sec:M2fluc-beta0} we expand the M2 brane action to  quadratic level and determine the spectrum of bosonic
and fermionic fluctuations.

 In section \ref{sec4} we consider the case of $\bb=0$ 
 and show how to reproduce the one-loop term in (\ref{1.10})
from the cusp anomaly expanded  in small  angle $\aa$.
This is achieved
by using (\ref{1.21}) and doing quantum-mechanical perturbation theory for the fluctuation energies in  small $\aa$.

Similar  analysis is repeated in section \ref{sec5} for the case of a cusped Wilson line
 with $\aa=0$  and small $\bb$  demonstrating that  this  leads to the same 
 expression \rf{1.22},\rf{1.23} for the \brem function, in agreement with the expected 
  BPS structure in (\ref{1.5}).

\section{String  in $\ads_{4}\times\cp$ and  M2 brane in \adsz  solutions representing  cusped Wilson line  }
\la{sec:classical}

\subsection{String  solution}

We shall  follow  \cite{Drukker:2011za,Forini:2012bb} 
(see also \cite{Drukker:2007qr}) 
and  use  global coordinates in $\ads_{4}$ as in \rf{1.11}. 
The $\cp$  metric  and $A$  in \rf{1.14}  can be expressed in terms of 6  real angles as 
 (see \eg  \cite{Bandres:2009kw})
\ba
ds_{\cp}^{2} =&{d}\gamma ^{2}+\cos ^{2}\gamma \sin
^{2}\gamma \Big( {d}\psi +\tfrac{1}{2}\cos \theta _{1}\, 
{d}\vp_{1}-\tfrac{1}{2}\cos \theta _{2}\, {d}\vp_{2}\Big) ^{2} \no 
 \\
&\ \ \ \ +\tfrac{1}{4}\cos ^{2}\gamma \left( {d}\theta _{1}^{2}+\sin ^{2}\theta
_{1}\, {d}\vp_{1}^{2}\right) +\tfrac{1}{4}\sin ^{2}\gamma \left(
{d}\theta _{2}^{2}+\sin ^{2}\theta _{2}\, {d}\vp_{2}^{2}\right)\ , \la{211}\\
A=&\tfrac{1}{2}\left( \cos 2\gamma \,d\psi + \cos ^{2}\gamma \cos \theta _{1} \,d\varphi _{1}+\sin ^{2}\gamma \cos \theta
_{2}\,d\varphi _{2} \right)\ , \la{5.5}
\ea
where $0\leq \gamma <{\pi \over 2},$\  $0 \leq \psi < 2\pi ,$\ 
 $0\leq \theta _{i}\leq \pi ,$ \  $0\leq \varphi _{i}<2\pi $.
We will  consider the configuration localised at 
\be\la{323}
\te  x=0\ , \qquad 
\gamma = \frac{\pi}{4}\ , \qquad \theta_{1,2}=\frac{\pi}{2}\ , \qquad \varphi_{1,2}=0\ , 
\ee
and embedded into 
 the   subspace   $\ads_{3}\times S^{1}\subset \ads_{4}\times \cp$ with the metric 
\be
\la{2.1}
ds^{2} = \tfrac{1}{4}R^{2}(-\cosh^{2}\rho\, dt^{2}+d\rho^{2}+\sinh^{2}\rho\, d\theta^{2})+\tfrac{1}{4} R^2d\psi^{2} \ . 
\ee
The IIA string solution corresponding to the cusped Wilson line 
   is described by 
\be
\la{2.2}
\rho=\rho(\theta)\ , \qquad\qquad  \psi=\psi(\theta)\ ,
\ee
with  $t$  and $ \theta$    identified with  the world-sheet coordinates. 
 We will also use  the  coordinates  $(\tau, \sigma)$, with $\tau$ proportional to $t$ and $\sigma$  being a particular  function of $\theta$. 
 
The  solution will  have the following range  of 
 $\theta(\sigma)$ and $ \psi(\sigma)$ 
\be\la{2.3}
\theta(\sigma)\in[\tfrac{\alpha}{2},\pi-\tfrac{\alpha}{2}]\ , \qquad \qquad \qquad 
\psi(\sigma)\in[-\tfrac{\beta}{2},\tfrac{\beta}{2}]\ ,
\ee
with the parameters $\alpha$ and $ \beta$ being  the spatial 
cusp angle  and the  internal cusp angle discussed in the Introduction. 
The coordinate $\rho(\sigma)$ will cover   the range $[\rho_{\rm min}, \infty]$ 
twice. 

The resulting induced string metric is 
\be
\la{2.4}
ds^{2} = \tfrac{1}{4}R^2 \big[-\cosh^{2}\rho\, dt^{2}+(\rho'^{2}+\sinh^{2}\rho+\psi'^{2})\, d\theta^{2}\big]\ .
\ee
The  classical Nambu-Goto   string action  is then given by 
\iffa 
\footnote{In more details,
$T_{2}\frac{2\pi R}{k}\frac{R^{2}}{4} = \T_{2}\frac{\pi}{2k} = \frac{\sqrt{2k N}}{\pi}\frac{\pi}{2k} =  \sqrt{\frac{\l}{2}}$
and is usually  written as $\sqrt{\bar\l}/(2\pi)$ with $\bar\l = 2\pi^{2}\l$. }
\fi
\be
\la{2.5}
S 
= T\int dtd\theta\, L \ , \qquad \qquad 
L= \cosh\rho\, \sqrt{\rho'^{2}+\sinh^{2}\rho+\psi'^{2}} \ , 
\ee
where the effective string  tension $T$ was  defined in \rf{1.199}.
From (\ref{2.5}) we get  the conserved quantities:   $\E$ (the ``energy"   conjugate to  $\theta$)  and $J$ (the momentum  corresponding to $\psi$).\footnote{Sign of $\E$ is conventional and we follow \cite{Drukker:2007qr}. The explicit expressions are 
$-\E = \rho'\frac{\partial L}{\partial\rho'}+\psi'\frac{\partial L}{\partial\psi'}-L =
 -\frac{\sinh^{2}\rho\cosh\rho}{\sqrt{\rho'^{2}+\psi'^{}+\sinh^{2}\rho}}$ and 
 $J=\frac{\partial L}{\partial\psi'} = \frac{\cosh\rho\ \psi'}{\sqrt{\rho'^{2}+\psi'^{}+\sinh^{2}\rho}}$.} 
The BPS limit is $\E = \pm J$ \cite{Drukker:2007qr}.\footnote{At these special points the classical action vanishes after adding 
a suitable boundary action required to have the correct Neumann boundary conditions for some of the string  coordinates \cite{Drukker:1999zq}.} It  is convenient to 
define the following constant parameters 
 \be\la{2.6}
p = \frac{1}{\E}\ ,\qquad\qquad  q = \frac{J}{\E}=\frac{\psi'}{\sinh^{2}\rho}\ .
\ee
Let us introduce the  function $\xi(\theta)$ defined  in terms of $\rho(\theta)$ by  \cite{Drukker:2011za}
\be
\la{2.7}
\xi(\theta) = \frac{1}{b}\sqrt\frac{p^{2}+b^{4}}{p^{2}\sinh^{2}\rho +b^{2}}\ , \qquad\qquad  b^{2} = \frac{p^{2}-q^{2}}{2}+\sqrt{p^{2}+\frac{(p^{2}-q^{2})^{2}}{4}}\ ,
\ee
and also the  parameter $\kk$  obeying 
\be\la{2.8}
\kk^{2} = \frac{b^{2}(b^{2}-p^{2})}{p^{2}+b^{4}}\ ,
\qquad \ \ \ 
p^{2} = \frac{b^{4}(1-\kk^{2})}{b^{2}+\kk^{2}}\ , \qquad\ \ \  q^{2} = \frac{b^{2}(1-2\,\kk^{2}-\kk^{2}b^{2})}{b^{2}+\kk^{2}}\ .
\ee
Solving the equations of motion 
for  $\theta(\xi)$ one finds that the cusp angles are given 
in terms of the elliptic functions (with modulus $\kk^2$)\footnote{One finds two branches for the functions $\theta(\rho)$ and $\psi(\rho)$: one with $\rho$ ranging  from $\infty$ (AdS boundary)
to a minimal value, and another  with $\rho$  growing from a minimal  value  to infinity.}
\be
\alpha = \pi-\frac{2p^{2}}{b\sqrt{p^{2}+b^{4}}}\Big[\Pi\Big(\frac{b^{4}}{p^{2}+b^{4}}\Big| \kk^{2}\Big)-\KK(\kk^{2})\Big]\ , \qquad\qquad 
\beta = \frac{2b q}{\sqrt{p^{2}+b^{4}}}\KK(\kk^{2})\ .\la{2.9}
\ee
The small angle limit is $p\gg 1$ with fixed $q$  when 
\ba
\la{2.11}
b &= p+\frac{1-q^{2}}{2p}+\cdots, \qquad \kk^2 = \frac{1-q^{2}}{p^{2}}+\cdots, \qquad
\alpha = \frac{\pi}{p}+\cdots, \qquad \beta = \frac{\pi q}{p}+\cdots\ .
\ea
The (regularized) value of the classical action  is 
\ba
S_{\rm cl} &= \TT\,  T\frac{2\sqrt{b^{4}+p^{2}}}{b p}\Big[\frac{(b^{2}+1)p^{2}}{p^{2}+b^{4}}\KK(\kk^{2})-\EE(\kk^{2})\Big] 
=  \TT\,  T\Big[\frac{\pi(q^{2}-1)}{2p^{2}}+\mc O(p^{-3})\Big] \no \\
&= \frac{1}{2\pi}\TT\, T (\beta^{2}-\alpha^{2})+\cdots,\qquad \qquad 
\TT\equiv \int dt \ . \la{2.12}
\ea
Comparing to \rf{1.5},\rf{1.19}  we conclude that we reproduce  the leading planar strong-coupling  (string  tree-level)  term  in the ABJM  \brem function   as given  in \rf{1.6}  
\be\la{2.122}
B^{(0)}=\frac{1}{2\pi} T=  \frac{1}{2\pi}\sqrt\frac{\l}{2}\ . 
\ee

\paragraph{Elliptic parametrisation} \ \ \

Replacing $\rho$ by  $\xi$ introduced in (\ref{2.7})
the induced metric  (\ref{2.4}) can  be written as 
\be\la{2.13}
ds^{2} =  \frac{R^2}{4\xi^{2}}\Big[
-\frac{1+b^{2}}{b^{2}}\frac{1-\kk^{2}\xi^{2}}{1-\kk^{2}}dt^{2}+\frac{d\xi^{2}}{1-\xi^{2}}\Big]\ , \\
\ee
Let us  now define the world-sheet coordinates $(\sigma,\tau)$ as 
\ba
\la{2.15}
\sigma =  F(\arcsin\xi|\kk^{2})-\KK\  & \in\  [-\KK,\KK]\ ,  \qquad \ \ \ \tau = \frac{1}{b}\sqrt\frac{1+b^{2}}{1-\kk^{2}}\ t\ \in\ \mathbb R\ ,  \\
\la{2.16}
&\xi = \sn(\sigma+\KK) = \frac{\cn\sigma}{\dn\sigma}  \ ,
\ea
where  $\KK\equiv  \KK(\kk^2)$, $F$ is the elliptic integral of the first kind
and $\cn(\sigma) \equiv \cn (\sigma|\kk^2)$ 
is a  Jacobi elliptic function.
This  allows to write the  metric \rf{2.13}  in the conformally flat form 
\be
\la{2.17}
ds^{2} =\tfrac{1}{4} R^2\frac{1-\kk^{2}}{\cn^{2}\sigma}(-d\tau^{2}+d\sigma^{2})\ .
\ee
Note that $\cn\s$ function here   has an  implicit dependence on $\kk$.
For zero cusp angle, i.e. $\kk=0$,  we get $\cn \s \to \cos \s$   so  that \rf{2.17}
reduces to the AdS$_2$  metric 
 \be
\la{2.177}
ds^{2} =\tfrac{1}{4} R^2\frac{1}{\cos^{2}\sigma}(-d\tau^{2}+d\sigma^{2})\ .
\ee
Expressed in terms of $\sigma$ the string  solution has the following  explicit form 
\be
\la{2.18}
\theta(\sigma) = \frac{\pi}{2}+\frac{p^{2}}{b\sqrt{p^{2}+b^{4}}}\Big[\sigma-\Pi(\tfrac{b^{4}}{p^{2}+b^{4}}, \text{am}(\sigma+\KK) | \kk^{2})
+\Pi(\tfrac{b^{4}}{p^{2}+b^{4}}|\kk^{2})\Big]\ , \qquad 
\psi(\sigma) = \frac{b q}{\sqrt{p^{2}+b^{4}}}\, \sigma\ .
\ee

\paragraph{Small angle expansions}\ \ \

Starting with the  relations in (\ref{2.7}),  (\ref{2.8}), (\ref{2.9})
and expanding at {large $p$ for  fixed $q$} gives
\ba
b^{2} &= p^{2}+1-q^{2}+\frac{q^{2}-1}{p^{2}}+\frac{(q^{2}-1)(q^{2}-2)}{p^{4}}+\cdots, \qquad 
\kk^{2} = \frac{1-q^{2}}{p^{2}}-\frac{(q^{2}-1)(q^{2}-3)}{p^{4}}+\cdots, \no \\
\la{A5}
\alpha &= \frac{\pi}{p}+\frac{\pi(3q^{2}-5)}{4p^{3}}+\cdots, \qquad\qquad 
\beta = \frac{\pi q}{p}+\frac{\pi q(q^{2}-3)}{4p^{3}}+\cdots.
\ea 
For   $\beta=0$ 
(that corresponds to $q=0$) 
the small  $\kk$ limit  is the same as the small $\alpha$  limit, i.e. we get 
\ba
\kk^{2} = \frac{1}{p^{2}}-\frac{3}{p^{4}}+\cdots,\qquad \alpha = \frac{\pi}{p}-\frac{5\pi}{4p^{3}}+\cdots,\qquad \qquad 
\la{a7}
\frac{\alpha^{2}}{\pi^{2}} = \kk^{2}+\frac{1}{2}\kk^{4}+\cdots.
\ea
Another useful limit is large  {$p$ at fixed imaginary $\kk$}. From the  relations in (\ref{2.8}) and (\ref{A5})  we find 
\ba\la{a77}
b^{2} &= \frac{p^{2}}{1-\kk^{2}}+\kk^{2}-\frac{(1-\kk^{2})\kk^{4}}{p^{2}}+\cdots, \qquad 
q^{2} = -\frac{\kk^{2}p^{2}}{1-\kk^{2}}+1-2\kk^{2}-\frac{\kk^{2}(1-2\kk^{2}+2\kk^{4})}{p^{2}}+\cdots, \no\\
&
\alpha = \mc O(p^{-1})\ , \qquad \qquad 
\frac{\beta^{2}}{\pi^{2}} = -\frac{(4-5\kk^{2})^{2}\kk^{2}}{16(1-\kk^{2})^{3}}+\mc O(p^{-2})\ . 
\ea
Thus the  case of $\alpha=0$ is obtained by taking $p\to \infty$. Then  further expansion in small $\kk$ corresponds to the small $\beta$ expansion (cf. \rf{a7})
\be\la{a8}
\frac{\beta^{2}}{\pi^{2}} = 
-\kk^{2}-\frac{1}{2}\kk^{4}+\cdots\ .  
\ee

\paragraph{Case of  $\beta=0$ } \ \ \ 

Let us  record the  explicit form of the solution in 
the special case of the  vanishing internal angle $\beta=0$   which  
is obtained for $q=0$. Then the above functions 
can be expressed in terms of the parameter $\kk$, \ie
\ba
\la{2.19}
&  b^{2} = \frac{1-2\kk^{2}}{\kk^{2}}, \qquad p^{2} = \frac{(1-2\kk^{2})^{2}}{\kk^{2}(1-\kk^{2})} \ , \qquad   \tau = \frac{1}{\sqrt{1-2\kk^{2}}}\, t  ,\ \ \ \ \ \ \cosh^{2}\rho = \frac{1-\kk^{2}}{1-2\kk^{2}}\frac{1}{\cn^{2}\sigma} \ ,  \\
\la{2.20}
&\theta(\sigma) = \frac{\pi}{2}+\kk\sqrt\frac{1-2\kk^{2}}{1-\kk^{2}}\Big[\sigma-\Pi(1-\kk^{2},\text{am}(\sigma+\KK)|\kk^{2})
+\Pi(1-\kk^{2}|\kk^{2})\Big]\ , \qquad \ \ \ \ 
\psi(\sigma) = 0\ .
\ea
In this case the relation between  the small cusp  angle $\alpha$ and $\kk\ll 1$ is given in \rf{a7} 
\be\la{200}
\a=\pi \big(\kk + \tfrac{1}{4} \kk^3 + ...\big) \ . 
\ee
For the subsequent analysis of fluctuations it is useful to record the  values of the following derivatives with respect to $\sigma$ 
\be
\la{2.24}
\theta'^2(\sigma) = \frac{\kk^{2}(1-\kk^{2})}{1-2\kk^{2}}\frac{1}{\sinh^{4}\rho}\ ,\qquad\qquad 
\rho'^{2}(\sigma) = \frac{-1+(1-2\kk^{2})^{2}\cosh^{2}(2\rho)}{4(1-2\kk^{2})\sinh^{2}\rho}\ .
\ee

 \subsection{M2 brane solution} 
 
It is straightforward  to    uplift of the above 10d string cusp solution to  11d
 M2  brane solution   wrapped also on 11d circle (cf. (\ref{1.11})-(\ref{1.14}))
 so that the latter reduces to the former upon the ``double dimensional reduction"
 \ci{Duff:1987bx}. 
Let us consider, for example, 
the special   case of $\beta=0$ when   the solution is
 localized at a point in $\CP^3$ (in addition to \rf{323}).
 Then 
the analog  of the  induced metric \rf{2.1}   
written in the world-volume  coordinates $(t,\theta, \phi)$   is 
\be
ds^{2} = g_{ij}d\sigma^{i}d\sigma^{j} = \tfrac{1}{4}R^2\big [-\cosh^{2}\rho\, dt^{2}+(\rho'^{2}+\sinh^{2}\rho)\,d\theta^{2}\big]
  +\frac{1}{k^{2}}R^{2} \,d\phi^{2} \ .
\ee
Note that as  follows from \rf{1.15}  
\ba
F_{4} &= dC_{3} = -\tfrac{3}{8}R^{3}\cosh\rho\, \sinh^{2}\rho\, \cos x\ dt\wedge d\rho\wedge d\theta\wedge dx\  , \\
\la{2.28}
C_{3} &= \tfrac{3}{8}R^{3}\cosh\rho\, \sinh^{2}\rho\, \sin x\, dt\wedge d\rho\wedge d\theta \ . 
\ea
Since on the solution $x=0$    we have  vanishing $C_{3}$  contribution to the 
M2  brane action in \rf{1.17}.

 The  volume part of the M2  brane  action is 
\be\la{2.244}
S_{\rm V}  =  \frac{1}{4k}\T_{2}\,\int dt\, d\theta\, d\phi\,\cosh\rho\, \sqrt{\rho'^{2}+\sinh^{2}\rho}\ ,
\ee
which is  the same   as in the type IIA string case, i.e.  
we again reproduce \rf{2.122}.\footnote{Since $\int d \phi=2\pi$   we get 
the  prefactor of the integral over $t$ and $\theta$ as 
$\frac{2\pi R}{k}\frac{R^{2}}{4} T_{2}= \frac{\pi}{2k}\T_{2} = \frac{\sqrt{2k N}}{\pi}\frac{\pi}{2k} =  \sqrt{\frac{\l}{2}}=T$.}

\section{M2 brane   fluctuations near AdS$_4$ cusp solution 
}
\la{sec:M2fluc-beta0}

Let us now derive   the  spectrum of  masses of quadratic fluctuations around the 
M2 brane analog of the $\beta=0$ solution (\ref{2.20}).
We shall use a static  gauge  where, in particular,  the  11d angle $\phi$ is identified with the 
second spatial  world-volume coordinate, i.e. is not fluctuating.

\subsection{Bosonic  fluctuations}

\paragraph{$S^{7}/\Z_{k}$ scalars}\ \ \ 

As the $\beta=0$ 
solution is  trivial in $\CP^3$,  the $S^{7}/\Z_{k}$  fluctuations
 are  completely decoupled from  the AdS$_4$  fluctuations. 
 This is similar to  the case of the AdS$_2\times S^1$ 
 M2  brane solution  corresponding to the  circular Wilson loop case 
in  \cite{Sakaguchi:2010dg,Giombi:2023vzu}.
 To quadratic order in the remaining  3  complex  fluctuations  of  $w^{s}$ in  \rf{1.14} (that have trivial classical values)  we have 
\ba
ds^{2}_{S^{7}/\Z_{k}} 
= dw^{s}d\wb^{s}+\frac{1}{k^{2}}d\phi^{2}+\frac{i}{k}(w^{s}d\wb^{s}-\wb^{s}dw^{s}) d\phi+\cdots.
\ea
Using the indices  $a,b = 0,1$ for  $(\tau, \sigma)$  and $i, j= 0,1,2$
for $(\tau, \sigma,\phi)$
and  denoting  by  prime the derivative with respect to $\phi$, we have (\cf  (\ref{2.17}))
\ba
ds^{2} &= g_{ij}d\sigma^{i}d\sigma^{j}\ , \qquad\qquad \ \ \ \  g_{ij} = g_{ij}^{(0)}+\delta g_{ij}\ , \\
\la{3.3}
g_{ij}^{(0)} &=  R^{2}\begin{pmatrix}
\frac{1}{4}\hat g_{ab} & 0\\
0  & \frac{1}{k^{2}}
\end{pmatrix}, \qquad \qquad \hat g_{ab}d\sigma^{a}d\sigma^{b} = \frac{1-\kk^{2}}{\cn^{2}\sigma}(-d\tau^{2} +d\sigma^{2}) \ , \\
\delta g_{ij} &= 
R^{2}\begin{pmatrix}
\partial_{a}w^{s}\partial_{b}\wb^{s} &
h_{a3}
\\  h_{b3} &
  w'^{s}\wb'^{s}+\frac{i}{k}(w^{s}\wb'^{s}-\wb'^{s}w^{s})
\end{pmatrix} \ , \la{3.4} \\
& h_{a3}=\te  \frac{1}{2}\wb'^{s}\partial_{a}w^{s}+\frac{1}{2}w'^{s}\partial_{a}\wb^{s}+\frac{i}{2k}(w^{s}\partial_{a}\wb^{s}-\wb^{s}\partial_{a}w^{s}) \ . 
\ea
\iffa
\delta g_{ij} &=  R^{2}\begin{pmatrix}
\partial_{a}w^{s}\partial_{b}\wb^{s} & \bullet\\
\bullet  & w'^{s}\wb'^{s}+\frac{i}{k}(w^{s}\wb'^{s}-\wb'^{s}w^{s})
\end{pmatrix} \ . \la{3.4}
\ea
\fi
The WZ part  of  the M2  brane action does not contribute  in  this sector (cf. \rf{2.28})  while 
the   quadratic  fluctuation term in the  volume part of the M2  brane 
action  is found to be 
\ba
 & T_{2}\int d^{3}\sigma\, \delta ( \sqrt{-g}\, )= \frac{1}{2}T_{2}\int d^{3}\sigma\, \sqrt{-g^{(0)}}g^{(0)ij}\delta g_{ij} \lp
= \frac{1}{8k}\T_2\int d\tau d\sigma d\phi\ \sqrt{-\hat g}\ \Big[4\hat g^{ab}\partial_{a}w^{s}\partial_{b}\wb^{s}
+k^{2}w'^{s}\wb'^{s}+ik(w^{s}\wb'^{s}-\wb'^{s}w^{s})\Big] \ . \la{35}
\ea
Expanding in Fourier modes  in the $\phi$ coordinate, $w^{s}(\tau,\sigma,\phi) = \sum_n e^{in\phi}w_{n}^{s}(\tau,\sigma)$,  we get for the corresponding  fluctuation Lagrangian 
\be  L_2 =4  \sqrt{-\hat g}\ \sum_n \Big[
\hat g^{ab}\partial_{a}w_{-n}^{s}\partial_{b}\wb_{n}^{s}+\tfrac{1}{4}(k^{2}n^{2}+2k n)\, w_{-n}^{s}\wb_{n}^{s}\Big] \ .\la{3.6}
\ee
It describes  three towers of complex scalar  2d  fluctuations propagating  in the metric $\hat g_{ab} $ in \rf{3.3}  with the   masses  (\cf (2.11) in \cite{Giombi:2023vzu})
\be
\la{3.8}  {S^{7}/\Z_{k}} :\qquad \ \ \ \ \  
m^{2}_n = \tfrac{1}{4}k n(k n+2) \ , \ \ \ \ \qquad    n=0, \pm 1, \pm 2, \dots\ .
\ee
The case of $n=0$  corresponds to the massless 2d fields  found in the string theory  case. 
For $k\ge 2$ the values of $m^2_n$  are positive.\foot{Note that for $ k=1$  and $n=-1$   we get 
$m^2_{-1} = -{1\ov 4}$ which saturates the stability bound  in AdS$_2$  
(to which the metric $\hat g_{ab}$ reduces in the zero cusp limit)
as the corresponding conformal dimension  given by 
$h = \ha(1+\sqrt{1+4m^2})  = \ha $.}

\paragraph{AdS$_4$  scalars}\ \ \ 

Using  the   AdS$_4 $   metric in 
\rf{1.12},\rf{1.13}  with the  coordinates $(t,\rho,x,\theta)$  
let us 
 fix the static gauge  as\foot{In the string  case an  alternative approach  is discussed  in  \ci{Drukker:2011za}.}
\ba \la{3.88}
&\delta t=0 \ , \qquad \ \ \  \delta \theta=0 \ , \qquad \ \ \ 
\delta \phi=0 \ , \ea
and define 
\ba
&\la{3.9} 
\delta \rho= - \sqrt{1 + \frac{\rho'^2}{\sinh^2\rho \ \theta'^2}  }
\  Z (\tau,\sigma,\phi)
\ , \qquad \qquad 
\delta x =   \frac{1}{\sinh\rho}\,  X(\tau,\sigma,\phi)
\ .  \ea
Here $\rho(\sigma)$ and $\theta(\sigma)$ 
 are the  classical solution functions 
 in (\ref{2.19}) and (\ref{2.20}) (see also  (\ref{2.24}))
 and $Z$ and $X$ represent   the two  non-trivial  AdS$_4$   3d fluctuation  functions. 
 Then  the quadratic term in the expansion  of the  volume term in the M2  action  may be written 
as
\be\la{3.11}
S_{2, \rm V} = \tfrac{1}{8} T_2 \int d\tau d\sigma d\phi\ \sqrt{-g^{(0)}} \, \Big[
R^2 (g^{(0)})^{ij}(\partial_{i}Z\partial_{j}Z+\partial_{i}X\partial_{j}X)
+{4}\big(4+R^{(2)}\big)\, Z^{2}+{8}X^{2}
\Big]\ . 
\ee
Here  $g^{(0)}_{ij}$ was defined in (\ref{3.3})   and  $R^{(2)}$ is the scalar curvature of  the 2d metric $\hat g_{ab}$ in \rf{3.3} 
\be
\la{3.14}
R^{(2)} = -2\Big(1+\frac{\kk^{2}}{1-\kk^{2}}\cn^{4}\sigma \Big)\ .
\ee
 Splitting the derivatives into 
 the  $\tau,\sigma$ and  $\phi$  ones   we   may write \rf{3.11} as 
\be\la{3.13}
S_{2, \rm V} = \tfrac{1}{8k}\T_{2}\, \int d\tau d\sigma d\phi\ \sqrt{-\hat g} \ \Big[
\hat g^{ab}(\partial_{a}Z\partial_{b}Z+\partial_{a}X\partial_{b}X)
+(4+R^{(2)})\, Z^{2}+2\,X^{2}+\tfrac{1}{4}k^{2}(Z'^{2}+X'^{2})
\Big]\ .
\ee
To find the   fluctuation part of the WZ term 
$S_{\rm WZ} = T_{2}\int C_{3}$ 
in the M2  brane action in \rf{1.17} we  note that according to  (\ref{2.28})
\ba
C_{3} &
= \tfrac{3}{8}R^{3}\cosh(\rho+\delta \rho)\sinh^{2}(\rho+\delta \rho)\, \frac{1}{\sinh\rho}(X+\cdots)\, \sqrt{1-2\kk^{2}}\, d\tau
\wedge (\rho'd\sigma+d\delta \rho)\wedge \theta'd\sigma\lp
=- \tfrac{3}{8}R^{3}\sqrt{1-2\kk^{2}}\cosh\rho\, \sinh\rho\  \theta'\,  X\,\delta\rho'\
d\tau\wedge d\sigma\wedge d\phi+\cdots.\la{3.133}
\ea
Note that  in our notation  $\rho'\equiv \del_\sigma \rho(\sigma)$, $\theta'\equiv \del_\sigma \theta'(\sigma)$, while for the  fluctuations 
$\delta\rho'(\tau,\sigma,\phi)  \equiv \partial_{\phi}\delta\rho$,  etc.
Expressing $\delta \rho$  in terms of  $Z$ in \rf{3.9} 
and  using the explicit  form of the solution  functions $\rho(\s),\theta(\s)$ we  get 
\be\la{3.144}
S_{2, \rm WZ} = \tfrac{3}{8}\T_{2}\, (1-2\kk^{2})\int d\tau d\sigma d\phi\ \cosh^{2}\rho\, XZ' = 
\tfrac{3}{8}\T_2\int d\tau d\sigma d\phi\ \sqrt{- \hat g}\,  XZ'
 \ .
\ee
Combining \rf{3.13}  and \rf{3.144} we get for the  corresponding quadratic fluctuation Lagrangian  (factoring out $\frac{1}{2}\sqrt{-\hat g}$)
\be\la{3.15}
L_{2,\rm AdS} =\hat g^{ab}(\partial_{a}Z\partial_{b}Z+\partial_{a}X\partial_{b}X) + (4+R^{(2)})\, Z^{2}+2\,X^{2}
+\tfrac{1}{4}k^{2}\, (Z'^{2}+X'^{2})+\tfrac{3}{2}k\, (XZ'-X'Z)\ .
\ee
Note that  for zero cusp angle, i.e. 
$\kk=0$,    when $\hat g_{ab}$ in \rf{3.3}  reduces to the AdS$_2$ metric
(cf. \rf{2.177})  and  $R^{(2)}$ in \rf{3.14}    takes the -2 value, 
the   mass terms  of $Z$ and $X$ become the same. Then    
  after the Fourier expansion in $\phi$  we get 
   the same two  towers  of massive  2d fields in AdS$_2$ 
   as  found in  \cite{Giombi:2023vzu}
\be
\la{3.20}
\kk=0:\ \ \ \ \qquad m^{2}_{n} = \tfrac{1}{4}(kn-2)(kn-4)\ .
\ee
 For generic $\kk$  expanding  in Fourier modes 
 one finds the following $4\times 4$  mass-squared matrix  for the 
 $\sin(n\phi)$, $\cos(n\phi)$  modes of $Z$ and $X$ 
\be\la{3.17}
{\cal M}^2_n=
\begin{pmatrix}
 2+\frac{R^{(2)}}{2}+\frac{k^2 n^2}{8} & 0 & 0 & \frac{3 k n}{4} \\
 0 & 2+\frac{R^{(2)}}{2}+\frac{k^2 n^2}{8} & -\frac{3 k n}{4} & 0 \\
 0 & -\frac{3 k n}{4} & 1+\frac{k^2 n^2}{8} & 0 \\
 \frac{3 k n}{4} & 0 & 0 & 1+\frac{k^2 n^2}{8} \\\end{pmatrix}\ .
\ee
The  corresponding eigenvalues are 
 (each with  multiplicity 2)
\be
\la{3.22}
m^{2}_{n,\pm} =3+\tfrac{1}{2} R^{(2)} +\tfrac{1}{4} k^{2}n^{2}\pm\tfrac{1}{2}\sqrt{(2+R^{(2)})^{2}+9k^{2}n^{2}}\ .
\ee
Considering the limit   when $\kk$ is small (and thus $2+R^{(2)}<0$ 
according to \rf{3.14})  we get\foot{The $\pm$  signs  in (\ref{3.22}) are  taken into account in \rf{3.24} 
 by the fact that $n\in\mathbb Z$.}    
\ba
& m^{2}_{0,\pm } =3+\tfrac{1}{2} R^{(2)}  \pm\tfrac{1}{2}|2+R^{(2)}|=
 \begin{cases}
2 \\
4+ R^{(2)} 
\end{cases}  =
 \begin{cases}
2 \\
2-2\cos^{4}\sigma\, \kk^{2}+\cdots
\end{cases},  \la{3.244}\\
\la{3.24}
 & m^{2}_{n,\pm} = \tfrac{1}{4}(kn-2)(kn-4)-\cos^{4}\sigma\, \kk^{2}+
  \cdots, \ \ \ \ \ \ \  n=\pm 1, \pm 2, \dots \ . 
\ea
Note that the  coefficient of the $\cos^{4}\sigma$ term  is different in the  cases  $n\neq 0$  and $n=0$.

\subsection{Fermionic fluctuations}

The general structure of the   fermionic part of the M2 action  in 11d background in \rf{1.17} is 
\ba
S_{F} &= T_{2}\int d^3\s\, \Big[\sqrt{-g}\, g^{ij}\partial_{i}X^{M}\bar\vartheta\Gamma_{M}\hat{D}_{j}\vartheta
-\tfrac{1}{2}\eps^{ijk}\partial_{i}X^{M}\partial_{j}X^{N}\bar\vartheta\Gamma_{MN}\hat{D}_{k}\vartheta+\cdots\Big]\ ,\la{3.222}\\
g_{ij}&=\partial_{i}X^{M}\partial_{j}X^{N}G_{MN}(X) \ ,\qquad \qquad G_{MN}=E^{A}_{M}E^{A}_{N}\ , \qquad\qquad  \Gamma_{M}=E^{A}_{M}(X)\Gamma_{A}\ , \\
\hat{D}_{i} &=\partial_{i}X^{M}\hat{D}_{M}\ , \qquad
\hat{D}_{M} = \partial_{M}+\tfrac{1}{4}\Gamma_{AB}\Omega^{AB}_{M}-\tfrac{1}{288}
(\Gamma\indices{^{PNKL}_{M}}-8\,\Gamma^{PNK}\delta^{L}_{M})\, F_{PNKL}\ .
\ea
In the type IIA GS  string limit  of the  cusp solution 
one finds  the  Dirac action for the 
3+3 2d fermions with   masses  $\pm 1$ and 
2 fermions    with   mass 0 \cite{Forini:2012bb}.
Being   independent of $\kk$  these   are the same  values  
 as in the  case of the  BPS  Wilson loop. 

In the  case of the AdS$_2\times S^1$  M2 brane solution without  zero cusp 
one finds as in   \cite{Sakaguchi:2010dg,Giombi:2023vzu}
 8  towers of    2d fermionic modes with  masses 
\be
\la{3.28}
m_n = \tfrac{1}{2}kn \pm 1\ \ \  (3+3\ \ \text{modes}\ \vartheta_\pm );  \qquad 
m_n= \tfrac{1}{2}kn\ \ \ (2\ \ \text{modes}\ \vartheta') \ , \qquad n=0, \pm 1, \pm 2, \dots\ .
\ee
These
reduce to the IIA  string values  for $n=0$.

 In the presence of the cusp, \ie for $\kk\neq 0$, 
the structure of the M2 brane action and the factorized   form of the  M2  brane solution in \adsz 
implies  that one can get the M2  brane fermionic masses from their IIA string values by the same 
  $\frac{1}{2}kn $ shift. The reason is that   the fermion operator
depends  just on the  induced metric and the $F_{4}$  background 
and  thus should be  universal.
A similar pattern in  the  structure of the 11d fermion spectrum  was observed in  \cite{Beccaria:2023ujc,Beccaria:2023sph,Beccaria:2023cuo}. 
Detailed form of   the Dirac operator in the induced metric (\ref{2.17}) will be given in the next section.

\section{One-loop M2  brane correction for  small cusp in AdS$_4$ 
}
\la{sec4}

The log of the  resulting one-loop M2 brane partition function \rf{1.19} may  be written as the sum   of the   contributions of 8+8 
bosonic   and fermionic   towers of 2d   fields, i.e. symbolically, 
\be \la{41}
- \log {\cal Z}_1=\ha   \sum_{n=-\infty}^\infty \sum_{p=1}^8 
\big( \log \det \Delta_{B_p,n} - 
 \log \det \Delta_{F_p,n} \big)\ . 
\ee
Here  the massive scalar Laplacian $\Delta_B$ and  the  fermionic  operator $\Delta_F$ are defined using the 2d   metric $\hat g_{ab}$ in \rf{3.3}
\be\la{4.1}
ds^{2} = \frac{1-\kk^{2}}{\cn^{2}\sigma}(-d\tau^{2}+d\sigma^{2})\ , \qquad \qquad -\KK<\sigma<\KK\ , 
\ee
i.e. $\Delta_B$    has the following structure
\be \la{43}
\Delta_B=- {1\ov \sqrt{-\hat g}}  \del_a (  \sqrt{-\hat g} \, \hat g^{ab} \del_b) + m^2
\ . \ee
Rescaling all the operators by $\sqrt{-\hat g}$  and using that \rf{4.1} is  conformally  flat   we get 
\be \la{430}
\Delta'_B= \del_\tau^2 - \del_\s^2    +\frac{1-\kk^{2}}{\cn^{2}\sigma}\, m^2 
\ . \ee
The  effect of  such  rescaling   should be trivial  after summing over $n$ 
 as there is no Weyl anomaly in 3 dimensions.\foot{The non-trivial  part of the 
 total Weyl anomaly cancels already  in the  GS  string case, i.e.  at the $n=0$ level  \ci{Drukker:2000ep}. 
 There is  formally  a residual UV   divergence proportional  to 
 the Euler number  that should  cancel against the  path integral measure
 \ci{Giombi:2020mhz}. This  divergence  and thus the Weyl anomaly 
  cancels automatically in the M2  brane case  as there are no log UV divergences in the 3d case \ci{Giombi:2023vzu}. Possible subtleties related to a
  Weyl rescaling of the 2d metric were discussed  
 in  the string   context   in \cite{Cagnazzo:2017sny}.
} 
Due to the translational invariance in $\tau$, integrating over the   corresponding 
momentum  component  one can, by the standard argument, express \rf{41} 
in terms of the   sum over the characteristic frequencies or eigenvalues of the 
spatial 1d operator in \rf{430}  (cf. \rf{1.20},\rf{1.21})
\be \la{44}
- \log {\cal Z}_1=\TT\,  E \ , \qquad \qquad   E=\tfrac{1}{2} \sum_{n=-\infty}^\infty \sum_{p=1}^8\sum_\ell  \Big[
\omega_{B_p,\ell}(n) - \omega_{F_p,\ell}(n) \Big] \ , \qquad \qquad   \TT=\int d \tau \ , \ee
where the  bosonic  frequencies $\omega_{\ell}$ are defined  by 
\be
\la{4.2}
\Big(-\partial_{\sigma}^{2}+ \frac{1-\kk^{2}}{\cn^{2}\sigma}m^{2}\Big)\Phi_{\ell}= \omega_{\ell}^{2 } \Phi_{\ell}\  .
\ee
The  fermionic  $\omega_{\ell}$   are  the eigenvalues of the corresponding Dirac operator 
\be
\la{4.3}
\Big[- i\,\gamma^{1}\,   \Big(\partial_{\sigma}+\frac{\sn\sigma\dn\sigma}{2\cn\sigma}\Big)
+\frac{\sqrt{1-\kk^{2}}}{\cn\sigma}\, m \Big]\Psi_{\ell}=\gamma^{0}\,  \omega_{\ell}
\Psi_{\ell} \ , 
\ee
where $\gamma^{a}$ are the  ``flat''  2d  gamma matrices defined  in terms of the  Pauli matrices $\ssigma_i$  as 
\be
\gamma^{a} = (\ssigma_{1},\ i\ssigma_{3})\ .
\ee
The explicit    values of the masses were given in 
 (\ref{3.8}),(\ref{3.22}) for  the bosons and in (\ref{3.28}) for the fermions. 

Given a complicated $\s$-dependent form of the operators in \rf{4.2},\rf{4.3} 
determining their spectrum is, in general,  a non-trivial problem. 
As we are interested in the correction to the \brem function, it is sufficient to find
the leading terms in $\omega_\ell$ in the small cusp angle $\alpha$ (or small $\kk$, cf. \rf{2.11}) expansion.
To do this we shall follow the same approach 
as was used  in 
 \cite{Aguilera-Damia:2014bqa} 
in the corresponding IIA  string case.\foot{This perturbative approach does not require to write the second order fluctuation
operators  in the explicit Lam\'e form
(for a similar near-AdS$_2$ expansion see   \cite{Forini:2017whz}).
 Also, as  discussed in \cite{Aguilera-Damia:2014bqa},  in this approach it is straightforward to 
implement the fermionic   boundary conditions compatible with the $\N=6$ supersymmetry in the zero cusp limit. 
} 

For $\kk=0$ we have $\cn \s \to \cos \s$   so that the operators in \rf{4.2},\rf{4.3} 
reduce to the corresponding  (rescaled) ones in the AdS$_2$ case. 
The associated  
 eigen-functions $\Phi_{\ell}, \Psi_{\ell}$ were  discussed, e.g., in \cite{Sakai:1984vm}.
For the scalars  with  the Dirichlet boundary conditions  one finds 
\ba
&\Phi_{h, \ell}(\sigma) = \frac{\sqrt{\ell!\, \Gamma(\ell+2h)\, (\ell+h)}}{2^{h-\frac{1}{2}}\, \Gamma(\ell+h+\frac{1}{2})}\cos^{h}\sigma\, 
P_{\ell}^{(h-\frac{1}{2}, h-\frac{1}{2})}(\sin\sigma)\ ,\la{49}
\\
\la{4.7}
&\ \ \  \int_{-\frac{\pi}{2}}^{\frac{\pi}{2}} \Phi_{h,\ell}(\sigma)\,\Phi_{h,\ell'}(\sigma)\, = \delta_{\ell,\ell'} \ , \\
\la{4.6}
&\qquad \omega_{\ell} =\ell+h\ , \qquad\qquad  h = \tfrac{1}{2}\big(1+\sqrt{1+4m^{2}}\big) \ ,  
\qquad \ \ \  \ell=0, 1, 2, \dots\ , 
\ea
where $h$   denotes the corresponding  \ads$_2$ conformal dimension (\ie $m^2 = h(h-1)$).
For the  fermions in $\ads_{2}$ with mass  $m<\ha$ we have 
\ba
\la{4.8}   \Psi_{h, \ell} = (\psi^{1}_{h,\ell}, \psi^{2}_{h, \ell}) \ , \qquad 
&\psi^{1}_{h, \ell}(\sigma) = \phantom{-} \frac{\sqrt{\ell!\, \Gamma(\ell+2h)}}{2^{h-\frac{1}{2}}\, \Gamma(\ell+h)}
\cos^{h}\sigma\cos(\tfrac{\sigma}{2}+\tfrac{\pi}{4})\, P_{\ell}^{(h, h-1)}(\sin\sigma)\ , \\
& \psi^{2}_{h, \ell}(\sigma) = -\frac{\sqrt{\ell!\, \Gamma(\ell+2h)}}{2^{h-\frac{1}{2}}\,\Gamma(\ell+h)}
\cos^{h}\sigma
\sin(\tfrac{\sigma}{2}+\tfrac{\pi}{4})\, P_{\ell}^{(h-1,h)}(\sin\sigma)\ ,\no  \\
&\omega_{\ell} =\ell+h \ ,\qquad  \qquad h=\tfrac{1}{2}-m \ .\la{498}
\ea
For the  fermions with $m>-\ha $ we have instead
\ba
\la{4.88}   \Psi_{h, \ell} = (\chi^{1}_{h,\ell}, \chi^{2}_{h, \ell}) \ , \qquad 
&\chi^{1}_{h, \ell}(\sigma) = \phantom{-} \frac{\sqrt{\ell!\, \Gamma(\ell+2h)}}{2^{h-\frac{1}{2}}\, \Gamma(\ell+h)}
\cos^{h}\sigma\cos(\tfrac{\sigma}{2}+\tfrac{\pi}{4})\, P_{\ell}^{(h-1, h)}(\sin\sigma)\ , \\
& \chi^{2}_{h, \ell}(\sigma) = -\frac{\sqrt{\ell!\, \Gamma(\ell+2h)}}{2^{h-\frac{1}{2}}\,\Gamma(\ell+h)}
\cos^{h}\sigma
\sin(\tfrac{\sigma}{2}+\tfrac{\pi}{4})\, P_{\ell}^{(h,h-1)}(\sin\sigma)\ ,\no  \\
&\omega_{\ell} =\ell+h \ ,\qquad  \qquad h=\tfrac{1}{2}+m \ .\la{499}
\ea
In both cases\foot{For fermions with $|m|\le \frac{1}{2}$ it is possible to adopt alternative quantizations, \ie choose one of the two possibilities above
\cite{Breitenlohner:1982jf,Breitenlohner:1982bm,Amsel:2008iz}.
Here this will happen only when $m=0$, which  may  occur (for generic  $k$) 
only  for  $n=0$.} 
\be
\int_{-\pi/2}^{\pi/2}\frac{d\sigma}{\cos\sigma}\Psi^{\dagger}_{h,\ell}\Psi_{h,\ell'} = \delta_{\ell,\ell'}\ .
\ee
Let us now find the first non-trivial terms in the  small $\kk$  expansion of the eigenvalues $\omega_\ell$ for   $n\neq 0$ first assuming  $k>2$
and then discussing  the  special   cases 
  of $k=1,2$. 

\subsection{Expansion of bosonic fluctuation  frequencies} 

\paragraph{$\cp$ scalars}\ \ \ 

For the 6 
 $\cp$  scalars  with the mass in (\ref{3.8}) the 
AdS$_2$ conformal dimension in \rf{4.6} is (assuming 
$k>2$)
\be
\la{4.12}
h_{n} = \tfrac{1}{2}+\tfrac{1}{2}|kn+1| = \begin{cases}
1+\frac{1}{2}kn\ , & n\ge 0\ , \\
-\frac{1}{2}kn\ , & n<0\ .
\end{cases}
\ee
As in  \cite{Aguilera-Damia:2014bqa} let us 
 rescale $\sigma\to \tilde \sigma $ and $\omega_{\ell}\to \tilde \omega_\ell $ as 
\be
\la{4.13}
\tilde\sigma = \frac{\pi}{2\KK}\sigma \in [-\frac{\pi}{2},\frac{\pi}{2}]\ , \qquad \tilde \omega_{\ell} = \frac{2\KK}{\pi}\omega_{\ell}\ , \qquad
\frac{2\KK}{\pi} = 1+\tfrac{1}{4}\kk^2+\cdots\ .
\ee
Then  (\ref{4.2}) takes the form 
\be\la{4.133}
\Big[-\partial_{\tilde \sigma}^{2}+(\frac{2\KK}{\pi})^{2}\frac{1-\kk^{2}}{\cn^{2}(\frac{2\KK}{\pi}\tilde \sigma)}m^{2} \Big]\Phi_{h,\ell}=\tilde\omega_{\ell}^{2} \Phi_{h,\ell} \ . 
\ee
As $m$ in \rf{3.8} is independent of $\kk$, expanding in small $\kk$ 
 gives
\be
\la{4.15}
\Big(-\partial_{ \tilde\sigma}^{2}+  \frac{m^{2}}{\cos^{2}\tilde\sigma}-\tfrac{1}{2}\kk^{2}m^2 +\cdots
\Big)\Phi_{h,\ell}=\tilde\omega_{\ell}^{2} \Phi_{h,\ell} \ , 
\ee
and thus using \rf{4.7} we get 
\be
\la{4.16}
\tilde\omega^{2}_{\ell} = (\ell+h_{n})^{2}-\tfrac{1}{2} \kk^{2}m^2 
\int_{-\pi/2}^{\pi/2}d\tilde\sigma\, \Phi_{h_{n},\ell}^{2} + \cdots
= (\ell+h_{n})^{2}-\tfrac{1}{2}\kk^{2}h_{n}(h_{n}-1)+\cdots\ . 
\ee
As a result, for the original $\omega_\ell$ in (\ref{4.13}) we obtain 
\be
\la{4.17}
\omega_{\ell} = (\ell+h_{n})(1-\tfrac{1}{4}\kk^2)-\frac{h_{n}(h_{n}-1)}{4(\ell+h_{n})}\kk^{2}+\cdots.
\ee

\paragraph{AdS$_4$ scalars} \ \ \

For the $\kk=0$   value of masses  in \rf{3.20}  $m^{2} = \frac{1}{4}(kn-2)(kn-4)$ we find from (\ref{4.6}) (assuming again  that  $k>2$)
\be
h_{n} = \tfrac{1}{2}+\tfrac{1}{2}|kn-3|
\la{4.19}
 = \begin{cases}
-1+\frac{1}{2}kn\ , & n>0\ , \\
2-\frac{1}{2}kn\ , & n\le 0\ .
\end{cases}
\ee
Taking into account  (\ref{3.24}), the expansion of (\ref{4.133}) reads
\be
\Big(-\partial_{ \tilde\sigma}^{2}+ \frac{h_{n}(h_{n}-1)}{\cos^{2}\tilde\sigma}-
\big[\tfrac{1}{2}h_{n}(h_{n}-1)+c_{n}\cos^{2}\tilde\sigma\big]\, \kk^{2}+\cdots
\Big)\Phi_{h_n, \ell}=\tilde\omega_{\ell}^{2}\Phi_{h_n, \ell} \ , 
\ee
where (cf. a  remark after (\ref{3.24}))
\be\la{cc}
c_{0}=2\ , \qquad\qquad  c_{n\neq 0}=1\ . 
\ee
Using that
\be
\la{4.22}
\int_{-\pi/2}^{\pi/2}d\tilde\sigma\, \Phi_{h,\ell}^{2}\cos^{2}\tilde\sigma = \frac{h(h-1)+(h+\ell)^{2}-1}{2(h+\ell+1)(h+\ell-1)}\ ,
\ee
we obtain 
\ba
\la{4.23}
\tilde\omega_{\ell}^{2} &= (\ell+h_{n})^{2}-\kk^{2}\int_{-\pi/2}^{\pi/2}d\tilde\sigma\, \Phi_{h_{n},\ell}^{2}\ \big[\tfrac{1}{2}h_{n}(h_{n}-1)+
c_{n}\cos^{2}\tilde\sigma\big]+\cdots \lp
= (\ell+h_{n})^{2}-\Big[\tfrac{1}{2}h_{n}(h_{n}-1)+c_{n}\frac{h_{n}(h_{n}-1)+(h_{n}+\ell)^{2}-1}{2(h_{n}+\ell+1)(h_{n}+\ell-1)}\Big]\, \kk^{2}+\cdots\ . 
\ea
For $n\neq 0$ we then get for $\omega_\ell$ in \rf{4.13}
\be
\la{4.24}
\omega_{\ell} = (\ell+h_{n})(1-\tfrac{1}{4}\kk^{2})-\Big[\frac{h_{n}(h_{n}-1)+1}{4(h_{n}+ \ell)}
+\frac{h_{n}(h_{n}-1)}{4(h_{n}+\ell)(h_{n}+\ell+1)(h_{n}+\ell-1)}
\Big]\kk^{2}+\cdots\, .
\ee
Note that the  sum over $\ell$ of the last term in square brackets is convergent  (and  independent of $h_{n}$)  
\be
\la{4.25}
\sum_{\ell=0}^{\infty}\frac{h_{n}(h_{n}-1)}{4(h_{n}+\ell)(h_{n}+\ell+1)(h_{n}+\ell-1)} = \frac{1}{8}\ .
\ee

\subsection{Expansion of fermionic fluctuation frequencies}

The values of the fermionic masses (that do not  depend on $\kk$)  were given in (\ref{3.28}). As was  mentioned above, 
one should consider separately the cases of  $m>-\ha$ and  $m<\ha$. Assuming $k>2$
we have 
\ba
n>0:\qquad \te \frac{kn}{2}\pm 1 > -\frac{1}{2}, \qquad \frac{kn}{2}>-\frac{1}{2}\ ; \qquad \qquad 
n<0:\qquad \te \frac{kn}{2}\pm 1 <\phantom{-} \frac{1}{2}, \qquad \frac{kn}{2}< \phantom{-} \frac{1}{2}.
\ea
Hence the corresponding values of $h_{n}$ and the spinor type (\ie with components $\psi$ as in (\ref{4.8})
or $\chi$ as in (\ref{4.88})) for the $3+3$ fermions 
 with mass $\ha kn\pm 1$ and $1+1$ fermions  with 
mass $\ha kn$ are as in Table \ref{tab:fermions}.
\begin{table}[H]
\be
\la{4.28}
\def\arraystretch{1.3}
\begin{array}{ccccc}
\toprule
\text{ } &
 m & \text{sign}(n) & h_{n} & \text{spinor }  \\
\midrule
\vartheta_{\pm} &
 \frac{kn}{2}\pm 1 & n>0 & \frac{kn}{2}\pm 1+\frac{1}{2} & (\chi^{1}, \chi^{2}) \\
\vartheta' & 
\frac{kn}{2} & n>0 & \frac{kn}{2}+\frac{1}{2} & (\chi^{1}, \chi^{2}) \\
\midrule
\vartheta_{\pm} & 
\frac{kn}{2}\pm 1 & n<0 & -\frac{kn}{2}\mp 1+\frac{1}{2} & (\psi^{1}, \psi^{2}) \\
\vartheta' & 
\frac{kn}{2} & n<0 & -\frac{kn}{2}+\frac{1}{2} & (\psi^{1}, \psi^{2}) \\
\bottomrule
\end{array}
\ee
\caption{
Mass, conformal dimension, and the spinor wave functions for the 8  towers of  fermionic modes. 
$\vartheta_{\pm}=(\vartheta_{\pm}^{r})$, $r=1,2,3$, represents 3+3 states, and $\vartheta'=(\vartheta'^s)$, $s=1,2$, 
represents 1+1  fermionic  states.
}
\la{tab:fermions}
\end{table}
After  the  rescaling in (\ref{4.13})  eq. (\ref{4.3})  reads
\be
\la{4.29}
\Big[-i\, \gamma^{1}\, \Big(\partial_{\tilde\sigma}+\frac{2\KK}{\pi}\frac{\sn\tilde\sigma\dn\tilde\sigma}{2\cn\tilde\sigma}\Big)
+\frac{2\KK}{\pi}\, \frac{\sqrt{1-\kk^{2}}}{\cn(\frac{2\KK}{\pi}\tilde\sigma)}\, m \Big]\Psi_{\ell}=\gamma^{0}\tilde\omega_{\ell}\Psi_{\ell} \  .
\ee
Expanding in small $\kk$ gives 
\be\la{4.31}
\Big[- i\, \gamma^{1} \Big(\partial_{\tilde\sigma}+\tfrac{1}{2}\tan\tilde\sigma+\tfrac{1}{8}\sin(2\tilde\sigma)\kk^{2}+\cdots\Big)
+\frac{m}{\cos\tilde\sigma}-\tfrac{1}{4}m\cos\tilde\sigma\, \kk^{2}+\cdots\Big]\Psi_{\ell}= \gamma^{0}
\tilde\omega_{\ell}\Psi_{\ell} \ . 
\ee
Next  we are to apply the standard  first-order perturbation theory 
using $\Psi_{h_{n},\ell}$  corresponding to a particular  value of $m$   in (\ref{4.28}).
We find\footnote{
Starting with a generic expression 
$
[i\, (\partial_{\sigma}+\frac{1}{2}\tan\sigma)\gamma^{1}+(\om+\kk^{2} \delta \om +\cdots)\gamma^{0}+B(\sigma)+C(\sigma)\kk^{2}+\cdots](\Psi+\delta\Psi\kk^{2}+\cdots)=0$, we get at first order 
$\int \frac{d\sigma}{\cos\sigma} \Psi^{\dagger}\gamma^{0}\big[i\, (\partial_{\sigma}+\frac{1}{2}\tan\sigma)\gamma^{1}+\om \gamma^{0}+B(\sigma)\big]\delta\Psi
+\int \frac{d\sigma}{\cos\sigma} \Psi^{\dagger}\big[\delta \om+\gamma^{0}C(\sigma)\big]\Psi$.
 Integrating by parts 
we find that  $\delta \om  = -\int d\sigma \Psi^{\dagger}\gamma^{0}C(\sigma) \Psi$.
}
\ba
\la{4.32}
\tilde\omega_{\ell} &= \ell+h_{n}-\kk^{2}\int_{-\pi/2}^{\pi/2}\frac{d\tilde\sigma}{\cos\tilde\sigma}\Psi_{h_{n},\ell}^{\dagger}\, \gamma^{0}
\Big[\tfrac{1}{8}\sin(2\tilde \sigma)\gamma^{1}+\tfrac{1}{4}m\cos\tilde\sigma\Big]\Psi_{h_{n},\ell}\lp
= \ell+h_{n}-\tfrac{1}{4}\kk^{2}\, m\int_{-\pi/2}^{\pi/2} {d\tilde\sigma\ov \cos \tilde \s}\ 
\Psi_{h_{n},\ell}^{\dagger}\gamma^{0}
\Psi_{h_{n},\ell}\ \cos\tilde\sigma\ ,
\ea
where we used that $\Psi^{\dagger}\gamma^{0}\gamma^{1}\Psi=0$ which is true for any $m>-\frac{1}{2}$ or $<\frac{1}{2}$. 
Here \ba
\int_{-\pi/2}^{\pi/2}\frac{d\tilde\sigma}{\cos\tilde\sigma}\Psi_{h,\ell}^{\dagger}\gamma^{0}
\Psi_{h,\ell}\ \cos\tilde\sigma &= \mp \frac{2(2h-1)(h+\ell)}{(2h+2\ell-1)(2h+2\ell+1)}
 \lp
= \mp\Big[\frac{2h-1}{2\ell+2h+1}+\frac{2h-1}{(2\ell+2h-1)(2\ell+2h+1)}\Big]\ 
\la{436} ,  \ea
where the 
{upper sign}  applies to the $(\psi^{1},\psi^{2})$ case 
and the   {lower sign}  to the $(\chi^{1},\chi^{2})$ case.
In \rf{436}   
we separated 
  a term which  gives  divergence when summed over $\ell$ and a term that gives a convergent 
$h$-independent sum
\be
\sum_{\ell=0}^{\infty}\frac{2h-1}{(2\ell+2h-1)(2\ell+2h+1)} = \frac{1}{2}\ .
\ee
Thus using  (\ref{4.28})    and (\ref{4.32}) we get  
\ba\la{4.34}
\omega_{\ell} = (\ell+h_{n})(1-\tfrac{1}{4}\kk^2 )\pm\tfrac{1}{4}m\Big[
\frac{2h_{n}-1}{2\ell+2h_{n}+1}+\frac{2h_{n}-1}{(2\ell+2h_{n}-1)(2\ell+2h_{n}+1)}
\Big]\,\kk^{2}+\cdots.
\ea
Using that $h=\frac{1}{2}\mp m$ (assuming  the  same  sign/spinor correspondence  as in \rf{436})   we end with 
\ba
\la{4.37}
\omega_{\ell} = (\ell+h_{n})\big(1-\tfrac{1}{4}\kk^{2}\big)+\tfrac{1}{4}\big(\tfrac{1}{2}-h_{n}\big)\Big[
\frac{h_{n}-\frac{1}{2}}{\ell+h_{n}+\frac{1}{2}}+\frac{h_{n}-\ha }{(\ell+h_{n}-\ha)(\ell+h_{n}+\ha)}
\Big]\,\kk^{2}+\cdots.
\ea
Notice that here there are no $\pm$ signs, \ie the expression is the same for 
both types
of the spinors once $m$ is written in terms of $h_{n}$.

\subsection{$n=0$: string theory limit  }

A similar  computation in the type IIA  string theory  case  was previously  done in \cite{Aguilera-Damia:2014bqa}. 
The string  case  corresponds to keeping 
only the  $n=0$  modes. To take this limit 
requires some care  as some details of the general $n$  formulae above depended 
on assumptions  that $n\neq 0$.
 For the  6   $\cp$ scalar fluctuations we can  set $h_0=1$  (\cf (\ref{4.12}))   and then from (\ref{4.17})
we  get  
\be
\la{4.38}
\omega_{\ell} = (\ell+1)(1-\tfrac{1}{4}\kk^{2})+\cdots.
\ee
For the two  AdS$_4$  scalar fluctuations we have the corresponding AdS$_2$   dimensions $h_{0}=2$ in (\ref{4.19}). One
mode  has  no corrections to its mass  in \rf{3.244}
 and  then 
\be
\la{4.39}
\omega_{\ell} = (\ell+2)(1-\tfrac{1}{4}\kk^{2})-\frac{1}{2(\ell+2)}\kk^{2}+\cdots.
\ee
For the other mode we need to use  $c_{0}=2$ in (\ref{4.23})   and this gives 
\ba
\tilde\omega_{\ell}^{2} &= (\ell+2)^{2}-\kk^{2}\int_{-\pi/2}^{\pi/2}d\tilde\sigma\, \Phi_{2,\ell}^{2}(1+2\cos^{2}\tilde\sigma)+\cdots = 
(\ell+2)^{2}-\frac{2(\ell+2)^{2}}{(\ell+1)(\ell+3)}\, \kk^{2}+\cdots\ ,  \la{4.40}
\ea
which leads to 
\ba
\la{4.41}
&\omega_{\ell}= (\ell+2)(1-\tfrac{1}{4}\kk^{2})-\Big[\frac{1}{\ell+1}-\frac{1}{(\ell+1)(\ell+3)}\Big] \kk^2 +\cdots.
\ea
For the  fermions, we have 1+1 states with $m=0$ and $h_0=\frac{1}{2}$ and 3+3 states with $m=\pm 1$ and $h_0=\frac{3}{2}$. 
Using (\ref{4.37}) we get 
\bea
\la{4.42}
m&=\ 0: \qquad \ \omega_{\ell} = (\ell+\tfrac{1}{2})(1-\tfrac{1}{4}\kk^{2})+\cdots, \\
m &=\pm 1: \qquad \omega_{\ell} = (\ell+\tfrac{3}{2})(1-\tfrac{1}{4}\kk^{2})-\Big[\frac{1}{4(\ell+1)}-\frac{1}{8(\ell+1)(\ell+2)} \Big] \kk^{2} + \cdots \ .
\eea

\subsection{One-loop  vacuum energy }

We are now ready to compute the total vacuum energy in \rf{44}
by summing the  characteristic frequencies $\om_\ell(n)$  over all modes.
We shall  first assume that $k>2$  and then consider the special  cases of $k=1,2$.

\paragraph{\ \ $\bm{k>2}:$}\ \ \ 
Let us  start  with the  $n>0$ modes  for which 
 (\cf (\ref{4.12}), \rf{4.19}, \rf{4.28})
\be
\la{4.43}
 \te h_{\cpp}=1+\frac{kn}{2}\ , \qquad 
h_{\aads}=\frac{kn}{2}-1\ , \qquad
h_{\vartheta_\pm} = \frac{kn}{2}\pm 1+\frac{1}{2}\ ,\qquad
h_{\vartheta'} = \frac{kn}{2}+\frac{1}{2}\ .
\ee
Doing  the sum over $\ell$  for each $n$  let us define  the contribution 
of each  of the 8+8  scalar and fermionic modes as 
\be\la{446}
 E_n =\tfrac{1}{2} \sum_{\ell=0}^\infty \om_{\ell}(n) \ . \ee
Expressing  the result in terms of 
 the Hurwitz zeta-function $\zeta(s,a) =\sum_{\ell=0}^\infty {1\ov (\ell +a)^s} $ 
we then get 
(including minus sign for the fermions)\foot{Note that the structure of the expressions  for $\vartheta_{\pm}$ and $\vartheta'$ fermions is the same due to the comment after (\ref{4.37}).}
\ba
E_n^{\rm CP} &= 6\times\Big[\tfrac{1}{2}\zeta(-1,h_{\cpp})(1-\tfrac{1}{4}\kk^{2})-\tfrac{1}{8}h_{\cpp}(h_{\cpp}-1)
\, \zeta(1,h_{\cpp})\kk^{2}+\cdots\Big]\ , \no\\
E_n^{\ads} &= 2\times\Big[\tfrac{1}{2}\zeta(-1,h_{\aads})(1-\tfrac{1}{4}\kk^{2})-\tfrac{1}{8}[h_{\aads}(h_{\aads}-1)+1]\, 
\zeta(1,h_{\aads})\, \kk^{2}-\tfrac{1}{16}\kk^{2}+\cdots\Big]\ , \no \\
E_n^{\vartheta_{\pm}} &= -3\times\Big[
\tfrac{1}{2}\zeta(-1,h_{\vartheta_{\pm}})(1-\tfrac{1}{4}\kk^{2})-\tfrac{1}{8}(\tfrac{1}{2}-h_{\vartheta_{\pm}})^{2}
\zeta(1,h_{\vartheta_{\pm}}+\tfrac{1}{2})\kk^{2}
+\tfrac{1}{16}(\tfrac{1}{2}-h_{\vartheta_{\pm}})\kk^{2}+\cdots
\Big]\ , \no \\ 
E_n^{\vartheta'} &= -2\times\Big[
\tfrac{1}{2}\zeta(-1,h_{\vartheta'})(1-\tfrac{1}{4}\kk^{2})-\tfrac{1}{8}(\tfrac{1}{2}-h_{\vartheta'})^{2}
\zeta(1,h_{\vartheta'}+\tfrac{1}{2})\kk^{2}
+\tfrac{1}{16}(\tfrac{1}{2}-h_{\vartheta'})\kk^{2}+\cdots
\Big]\ .\la{448}
\ea
Using  the  explicit values of dimensions $h_n$ in (\ref{4.43}) we get 
for the total 
\ba
E_n =& E_n^{\rm CP} + E_n^{\ads}+ 
E_n^{\vartheta_{+}}+ E_n^{\vartheta_{-}}+ E_n^{\vartheta'}\no  \\
 =& \zeta (-1,-1+\tfrac{k n}{2})-\tfrac{3}{2} 
\zeta(-1,-\tfrac{1}{2}+\tfrac{k n}{2})-\zeta (-1,\tfrac{1}{2}+\tfrac{k 
n}{2})+3 \zeta (-1,1+\tfrac{k n}{2})-\tfrac{3}{2} \zeta (-1,\tfrac{3}{2}+\tfrac{k n}{2})\lp
+\Big[\tfrac{1}{8} (-1+2 k n)+\tfrac{3}{32} 
(-2+k n)^2 {\zeta (1,\tfrac{k n}{2})}+\tfrac{1}{16} (-12+6 k n-k^2 n^2) 
{\zeta (1,-1+\tfrac{k n}{2})}\lp
\ \ \ -\tfrac{1}{8} k n (3+k n) {\zeta (1,1+\tfrac{k 
n}{2})}+\tfrac{3}{32} (2+k n)^2 {\zeta (1,2+\tfrac{k n}{2})}-\tfrac{1}{4} 
\zeta (-1,-1+\tfrac{k n}{2})\la{449}\\
&\ \ \ +\tfrac{3}{8} \zeta 
(-1,-\tfrac{1}{2}+\tfrac{k n}{2})+\tfrac{1}{4} \zeta 
(-1,\tfrac{1}{2}+\tfrac{k n}{2})-\tfrac{3}{4} \zeta (-1,1+\tfrac{k 
n}{2})+\tfrac{3}{8} \zeta (-1,\tfrac{3}{2}+\tfrac{k n}{2})\Big]\kk^{2}+\cdots.\no
\ea
This expression can be simplified   using the relations 
\ba
&\te  \zeta(1,-1+\tfrac{kn}{2}) = \zeta(1,\tfrac{kn}{2})+\frac{1}{\frac{kn}{2}-1}\ , \qquad \qquad  
\zeta(1,1+\tfrac{kn}{2}) = \zeta(1,\tfrac{kn}{2}) -\frac{4}{kn}\ , \\
&\te
\zeta(1,2+\tfrac{kn}{2}) = \zeta(1,\tfrac{kn}{2}) -\frac{4}{kn}-\frac{1}{1+\frac{kn}{2}}\ ,\qquad \zeta(-1,a) = \tfrac{1}{12}(-1+6a-6a^{2})\ . \la{450}
\ea
One  finds that all  log divergent  terms 
proportional to  $\zeta(1, \tfrac{kn}{2})$  cancel out. Also, the 
$\kk$-independent terms  combine to zero (reflecting the vanishing of vacuum energy in the AdS$_2$ limit \ci{Drukker:2000ep}) so that
\be\la{451}
E_{n} = \frac{6-5kn}{4kn(kn-2)}\ \kk^{2}+\mc O(\kk^4)\ . 
\ee
Let us  note that in \rf{449}  we used  the zeta-function regularization 
to subtract  the   linear divergences present  in the sum over $\ell$ for the  contributions of the 
individual fluctuations. In fact, this regularization is not
 necessary
once all these contributions are combined together -- the  resulting sum over $\ell$ is manifestly finite.  Indeed, 
the explicit form of the total 
 $E_n$ as the sum over $\ell$   is given by 
\ba
&E_n =  \Big[\tfrac{1}{8}(2kn-1)+\sum_{\ell=0}^{\infty} e_{n,\ell}\Big]\  \kk^2  +\mc O(\kk^4) \ ,  \la{99}  \\
& e_{n,\ell}=  -\frac{1}{2}-\frac{1+l+l^2}{2 (-2+2 l+k n)}+\frac{3 (1+l)^2}{4 (2 l+k 
n)}-\frac{(1+l) (-1+2 l)}{2 (2+2 l+k n)}+\frac{3 (1+l)^2}{4 (4+2 l+k 
n)} \ . \la{999}
\ea
Here $e_{n,\ell}\big|_{\ell \gg 1} = -\frac{1}{8}(12 + k^2 n^2) \ell^{-2} + \mc O(\ell^{-3})$
so that   the sum  over $\ell$ is convergent  and is given by 
\be \sum_{\ell=0}^{\infty}e_{n,\ell} = \frac{12-12 k n+5 k^2 n^2-2 k^3 n^3 }{8 k n (k n-2)}
\ . \ee 
As a result, \rf{99}  gives  the same result as in (\ref{451})
without using the zeta-function regularization.



One  can check that the  same expression is found also  for $n <0$ when 
\be\te
 h_{\cpp}=-\frac{kn}{2}\ , \qquad 
h_{\aads}=2-\frac{kn}{2}\ , \qquad
h_{\vartheta_\pm} = -\frac{kn}{2}\mp 1+\frac{1}{2}\ ,\qquad
h_{\vartheta'} = -\frac{kn}{2}+\frac{1}{2}\ .
\ee
Summing over $n\neq 0$ then gives 
\ba
\la{4.54}
\sum_{n\neq 0}E_{n} &= \sum_{n=1}^{\infty}\Big[
\frac{6-5kn}{4kn(kn-2)}+ \frac{6+5kn}{4kn(kn+2)}\Big]\kk^{2}+\cdots
= -2\sum_{n=1}^{\infty}\frac{1}{k^{2}n^{2}-4}\,\kk^{2}+\cdots\lp
= \Big(-\frac{1}{4}+\frac{\pi}{2k}\cot\frac{2\pi}{k}\Big)\, \kk^{2}+\cdots.
\ea
Note that this   sum  over $n$  is  also manifestly convergent. This   is to be 
compared to the 
Wilson loop case 
 in \cite{Giombi:2023vzu}  where  the sum  over the  M2  brane   modes in the  partition function  was  linearly divergent  and was defined using 
 the Riemann   zeta-function regularisation.\foot{Explicitly,   
 in \cite{Giombi:2023vzu} one had for the one-loop M2  brane correction  
  to $-\log \vev{ W }$: 
 $\Gamma_1= \sum_{n=1}^\infty \log \big( {k^2 n^2 \ov 4} -1 \big)$. This  gave
 $\Gamma_1=\log\big[2 \sin {2\pi\ov k} \big]$ 
 after using that $\zeta(0) = -\ha$, \ $ \zeta'(0) = - \ha  \log (2 \pi)$.}

It remains  to add the contribution of the 
string-level $n=0$ fluctuations that  can be  computed   using  \rf{4.38},\rf{4.39},\rf{4.41},\rf{4.42}) as was already done  in  \cite{Aguilera-Damia:2014bqa}.
  Explicitly, we  find
\be\la{455}
E_{0}=\sum_{\ell=0}^{\infty}\frac{1}{4(\ell+1)(\ell+2)}\ \kk^{2}+ \mc O(\kk^4)
=  \frac{1}{4}\kk^{2} + \mc O(\kk^4)\ .
\ee
Here the   sum over $\ell$ is also convergent, i.e. does not require to use the 
zeta-function regularization. 

Combing \rf{4.54} and \rf{455} we  conclude that 
\be \la{456}
E= \sum_{n=-\infty}^\infty E_n=
 \frac{\pi}{2k}\cot\frac{2\pi}{k}\ \kk^{2}+\mc O(\kk^4) \ . 
\ee
Using that in the small cusp angle limit  $\a= \pi \kk + ...$ (see \rf{200})
we observe  that  the one-loop  correction 
$\Gamma^{(1)}_{\rm cusp} = E$  in \rf{1.20}  indeed scales as $\alpha^2$. 
Then  using \rf{1.5}   we conclude that the one-loop correction to the \brem function  is given by \rf{1.22}.


\paragraph{\ \ $\bm{k=1}:$}\ \ \ 
For $k=1$ the relation  (\ref{4.12})  for $h_n$ for the  $\cp$   scalars   still applies. 
Eq. (\ref{4.19}) for $h_n$ for AdS$_4$ scalars  applies for $n>3$  while 
 for $n=1,2$  one has 
\be\te
h_{1,\aads} = \frac{1}{2}+\frac{1}{2}|1-3| =\frac{3}{2}\ ,\qquad  \qquad 
h_{2,\aads} = \frac{1}{2}+\frac{1}{2}|2-3| = 1\ .
\ee
For the fermions, the data in Table (\ref{4.28}) requires a modification for the mass $m_n=\frac{kn}{2}\mp 1$ for $n=\pm 1$. In these cases one has
\ba\te
m_{1} = -\frac{1}{2}\quad\to\quad h_{1}^{\vartheta_{-}} = 1\ ;\qquad  \qquad
m_{-1} = \frac{1}{2}\quad\to\quad h_{-1}^{\vartheta_{+}} = 1\ .
\ea
With these changes we find    
\be
 \te E_{1}=-\frac{1}{4}\kk^{2}+\cdots\ ,\qquad E_{2}=-\frac{5}{8}\kk^{2}+\cdots\ , 
\qquad E_{-1}= \frac{11}{12}\kk^{2}+\cdots\ .
\ee
Using also that for $k=1$ we have 
$E_{-2} = \frac{1}{2}$
the analog of the  sum in (\ref{4.54}) is  found to be 
\be\la{459}
\sum^\infty_{n=-\infty}E_{n}={ {\frac{1}{4}\kk^{2}}
+{(-\frac{1}{4}-\frac{5}{8}+\frac{11}{12}+\frac{1}{2})\kk^{2}}}
+ \sum_{n=3}^{\infty}\Big[
\frac{6-5n}{4n(n-2)}+ \frac{6+5n}{4n(n+2)}\Big]\kk^{2}+\cdots = -\frac{1}{4}\kk^{2}+\cdots.
\ee

\paragraph{\ \ $\bm{k=2}:$} \ \ \ 
For $k=2$,  (\ref{4.12}) is still valid while in 
 (\ref{4.19})  for $n=1$ one should  use 
$
h_{1} = \frac{1}{2}+\frac{1}{2}|2-3| =1$. 
For the fermions, the  values  in Table (\ref{4.28}) apply also for  $k=2$.
As a result,  $
E_{1} =-\frac{5}{8}\kk^{2}+\cdots$.
Using also that $E_{-1} =\frac{1}{2} $  we  find 
\be
\sum^\infty_{n=-\infty}E_{n} = {\frac{1}{4}\kk^{2}}
+{(-\frac{5}{8}+\frac{1}{2})\kk^{2}}
+ \sum_{n=2}^{\infty}\Big[
\frac{6-10n}{8n(2n-2)}+ \frac{6+10n}{8n(2n+2)}\Big]\kk^{2}+\cdots = -\frac{1}{4}\kk^{2}+\cdots.\la{461}
\ee
As as result, from  \rf{459} and \rf{461} we get  the corresponding values
  \rf{1.23} for the  one-loop correction to the \brem function.

\section{One-loop M2  brane correction for  small cusp in $\cp$ 
}
\la{sec5}

Let us now discuss the case   of  the  ``internal'' cusp, \ie  
when  $\alpha=0$   while  $\beta$ is non-zero.
Considering the small $\beta$ limit we will confirm that the coefficient of the leading 
$\beta^2$  in the corresponding  one-loop correction \rf{1.20},\rf{1.21}
is indeed  the same up to sign  as of the $\a^2$ term computed above, i.e. 
is   given again by \rf{1.22},\rf{1.23}.\foot{Let us note that  this case 
 has a relation to  the  latitude Wilson loop
for which there is no so far an exact localization result  for the expectation  value for  a
finite   angle.} 

The  $\alpha=0$  case  corresponds to  $p\to \infty$ at fixed imaginary $\kk$  when (see \rf{a77},\rf{a8},\rf{2.15}) 
\ba
& b
= \frac{p}{\sqrt{1-\kk^{2}}}\to \infty  \ , 
\qquad\qquad 
 t = {\sqrt{1-\kk^{2}}}\,\tau  ,\qquad \theta'(\sigma)=0 \ ,  \qquad \kk=i \epsilon\ , \la{51}  \\
& \psi'^{2} = -\kk^{2}, \qquad \qquad 
\rho'^{2} =1+(1-\kk^{2}) \sinh^{2}\rho \qquad \  
 \cosh^{2}\rho = \frac{1}{\cn^{2}\sigma}\ , \la{52}\\
 & {\beta^{2}}= -\pi^2 \frac{(4-5\kk^{2})^{2}\kk^{2}}{16(1-\kk^{2})^{3}}\ . \la{53}
\ea
The small $\beta$ limit   corresponds to $\kk\ll 1  $  when   
$\beta = i \pi \kk + ...$   which is  consistent with  the   range of $\psi= i\kk \sigma$  being $[-\ha \beta, \ha \beta]$  as in \rf{2.3}. 


\subsection{Fluctuation spectrum}

Let us  start with bosonic fluctuations  and  fix a static gauge similarly to  \rf{3.88}  
(cf. \cite{Drukker:2011za})
\ba \la{3.889}
&\delta t=0 \ , \qquad \ \ \  \delta \psi=0 \ , \qquad \ \ \ 
\delta \phi=0 \ , \\
&\la{5.7} 
\delta x =   \frac{1}{\sinh\rho}\,  X(\tau,\sigma,\phi)\ , \qquad
\delta \rho=- {  i \ov \kk}  \sqrt{\rho'^2  -\kk^2}  
\  W (\tau,\sigma,\phi)
\ , \qquad 
\delta \theta =   \frac{1}{\sinh\rho}\,  V(\tau,\sigma,\phi)
\ .  \ea
The ``volume'' part of the 
 quadratic fluctuation action then takes the form (cf. \rf{35},\rf{3.11})\foot{We use 
 tilde to denote fluctuations of the  $\cp$ angles  that have fixed values in \rf{323}
 and rescale $\tilde\gamma\to \frac{1}{2}\tilde \gamma$.}
\ba
\la{5.8}
S_{2, \rm V} =& \tfrac{1}{4}R^2 T_2\int d\tau d\sigma d\phi\ \sqrt{-g^{(0)}}\Big\{ 
\tfrac{1}{2} \Big[(g^{(0)})^{ij}(\partial_{i}V\partial_{j}V+\partial_{i}W\partial_{j}W+\partial_{i}X\partial_{j}X)\lp \te
+m^2_{V}\,V^{2}
+m^2_{W}\,W^{2}
+m^2_{X}\,X^{2}\Big]
{-\frac{4k}{R^{2}} \sqrt{\frac{\kk^{2}\cn^{2}\sigma}{1-\kk^{2}} +1 }\, \tilde\gamma\, \partial_{\phi}W} \\
&\te
+(g^{(0)})^{ij}(
2\,\partial_{i}\tilde \gamma\partial_{j}\tilde \gamma
+\frac{1}{4}\,\partial_{i}\tilde \theta_{r}\partial_{j}\tilde \theta_{r}
+\frac{1}{4}\,\partial_{i}\tilde \vp_r\partial_{j}\tilde \vp_r
)+2m^2_{\gamma}\tilde\gamma^{2}
-\frac{k}{R^{2}}\tilde\theta_{r}\partial_{\phi}\tilde\vp_r
-{i\ov 2 \kk} m^2_\gamma\, 
 \tilde\theta_{r}\partial_{\sigma}\tilde\vp_r
\Big\}\, , \no\\
m^2_{V} &\te = m^2_{X} = \frac{8}{R^{2}} +  m^2_\gamma\ , 
\qquad
m^2_{W} = \frac{4}{R^{2}}\big(2+R^{(2)}) + m^2_\gamma\ , 
 \qquad
m^2_{\gamma} \te = \frac{4}{R^{2}}\frac{\kk^{2}}{1-\kk^{2}}\cn^{2}\sigma\ .\no
\ea
Here  the metric  $g^{(0)}_{ij}$   is the same as in \rf{3.3}
and the  $\cn \s$ factors may be written in terms of the  conformal factor of the $\hat g_{ab}$   metric  as $\sqrt{-\hat g} = { 1-\kk^2\ov \cn^2\s}$. 
The two   pairs of fields $(\tilde\theta_{r},\tilde \varphi_{r})$ ($r=1,2$)
enter symmetrically   and   to diagonalize their Lagrangian we may 
perform the redefinition 
\be
\la{5.12}\te
\tilde \theta_{r} = \sqrt 2\,  ( A_r\, \cosh\frac{k\sigma}{2}-i  B_r\, \sinh\frac{k\sigma}{2})\ , \qquad\qquad
\tilde \varphi_r =  \sqrt 2\,  (B_r\, \cosh\frac{k\sigma}{2}+i  A_r\, \sinh\frac{k\sigma}{2})\ , 
\ee
 and use   \rf{52}. 
 \iffa 
  to get 
their Lagrangian as 
\be L_{A,B}=\te  \ha (g^{(0)})^{ij} ( \del_i A_r \del_j A_r + \del_i B_r \del_j B_r )
+ \ha m^2_\cpp ( A_r^2 + B_r^2 )\ , \qquad 
m^2_{\cpp} = \frac{\kk^2}{R^{2}}\frac{\kk^{2}}{1-\kk^{2}}\cn^{2}\sigma= \ . 
\la{5.15} \ee
\fi
As a result, we  can put \rf{5.8} into the form\footnote{The reason why the simple rotation in (\ref{5.12}) works is  due to the special dependence of the $(\tilde \theta, \tilde \vp)$ 
  mixing term in \rf{5.8} on  $\sigma$.
  Indeed, if we start with a model Lagrangian 
$L = \sqrt{-\hat g}\, \hat g^{ab}\big[ \partial_{a} \Phi_1\partial_{b}\Phi_1+\partial_{a}\Phi_2\partial_{b}\Phi_2
+\mu(\sigma) \Phi_1\partial_{\sigma}\Phi_2\big]$
 and assume that 
 $\mu(\sigma) = \frac{c}{\sqrt{-\hat g}} $
 where   $c$ is  a  constant   then  for a 
 conformally flat $\hat g_{ab}$   its   conformal 
 factor drops out, i.e. the  corresponding action is  the same as  in flat space. Thus  
it 
 can be diagonalized by  a rotation with an angle $\frac{c}{2}\sigma$. 
The resulting  mass  term    is then proportional to 
$1\ov \sqrt{-\hat g}$  like  $m_{A}^{2}$ in (\ref{5.17}).
}
\ba
& \no  S_{2, \rm V} =
\tfrac{1}{8}R^2 T_2\int d\tau d\sigma d\phi\ \sqrt{-g^{(0)}}\Big[ (g^{(0)})^{ij}\Big(\partial_{i}V\partial_{j}V+\partial_{i}W\partial_{j}W+\partial_{i}X\partial_{j}X\\ &
+\partial_{i}\tilde \gamma\partial_{j}\tilde \gamma
+\partial_{i}A_r\partial_{j}A_r
+\partial_{i}B_r\partial_{j}B_r\Big) +m^2_{V}\,V^{2}
+m^2_{W}\,W^{2}
+m^2_{X}\,X^{2} + m^2_{\gamma}\tilde\gamma^{2}+m_{A}^{2}(A_r^{2}+B_r^{2})
\lp
-\tfrac{2k}{R^2}
{\te  \sqrt{ 1 + R^2 m^2_A }  }\, 
(\tilde\gamma\, \partial_{\phi}W-W \partial_{\phi}\tilde\gamma) 
-\tfrac{2k}{R^{2}}(A_r\partial_{\phi}B_r-B_r\partial_{\phi}A_r)
\Big]\ ,\la{5.17}\\
 & \qquad m^2_{A} =\tfrac{1}{4} m^2_\gamma=  \tfrac{1}{R^{2}}\tfrac{\kk ^{2}}{1-\kk^{2}}\cn^2 \sigma 
= \tfrac{\kk^2}{R^{2}} \tfrac{1}{ \sqrt {-\hat g}}\ , \no 
\ea
where we used that $g^{(0)}_{ij}$ is given by  \rf{3.3}.
 In the IIA string limit
when terms  with $\del_\phi$ are absent  this action for 8 bosonic fluctuations  
is equivalent  the one found in  \cite{Forini:2012bb}. 

The contribution  of the WZ term  is  found using \rf{2.28} as in \rf{3.133}.
As here $\theta'=0$ (see \rf{51}) we get 
\ba
C_{3} &
= \tfrac{3}{8}R^{3}\cosh(\rho+\delta \rho)\sinh^{2}(\rho+\delta \rho)\, \frac{1}{\sinh\rho}(X+\cdots)\, \sqrt{1-\kk^{2}}d\tau
\wedge (\rho'd\sigma+d\delta\rho)\wedge d\delta\theta\lp
= \tfrac{3}{8}R^{3}\sqrt{1-\kk^{2}}\cosh\rho\, \sinh\rho\, X\, \rho'(\delta\theta)'\
d\tau\wedge d\sigma\wedge d\phi+\cdots, 
\ea
where again  $\rho'\equiv \del_\sigma \rho(\sigma)$ while the  prime on  all  the 3d fluctuation fields  is assumed to be the derivative over $\phi$, i.e. 
$(\delta\rho)' \equiv \partial_{\phi}\delta\rho$, $(\delta\theta)' \equiv \partial_{\phi}\delta \theta$.
We may  now use \rf{52}  and  (\ref{5.7})  to get 
\ba
\la{5.19}
S_{2, \rm WZ} &= \tfrac{3}{8}T_{2}R^{3}\sqrt{1-\kk^{2}}\int d\tau d\sigma d\phi\ \cosh\rho\, \rho'\, X\, V' 
= \tfrac{3}{8} {\T_{2}}\int d\tau d\sigma d\phi\ \sqrt{- \hat g}\  
{\te  \sqrt{ 1 + R^2 m^2_A }  }
\ 
 XV' \ . 
\ea
According to  \rf{3.3},  $\sqrt{-g^{(0)}} = \frac{R}{k}\sqrt{-\hat g}$  so we get from  (\ref{5.17})
 and  (\ref{5.19}) 
\ba
S_{2, \rm V} + S_{2, \rm WZ}  &
= \tfrac{1}{8k}\T_{2} \int d\tau d\sigma d\phi\ \sqrt{-\hat g}\ \Big( L_{\rm kin}+L_{\rm mass}+L_{\rm mix}\Big)\ ,\la{520}  \\
L_{\rm kin} &= \hat g^{ab} \Big(\partial_aV\partial_bV+\partial_aX\partial_bX
+\partial_aW\partial_bW
+\partial_a\tilde \gamma\partial_b\tilde \gamma
+\partial_aA_r\partial_bA_r
+\partial_aB_r\partial_bB_r\Big)\ ,\\
L_{\rm mass} &=\tfrac{R^{2}}{4}\Big[m^2_{V}\,V^{2}
+m^2_{X}\,X^{2}+m^2_{W}\,W^{2}+m^2_{\gamma}\tilde\gamma^{2}+ m^2_A (A_r^{2}+B_r^{2})\Big]
\lp \ \ \ \ \ \ \ \ 
+\tfrac{k^{2}}{4}\Big(V'^{2}+W'^{2}+X'^{2}+\tilde\gamma'^{2}+A_r'^{2}+B_r'^{2}\Big)
\no \\
& \qquad  - \tfrac{k}{2} \Big( {\te  \sqrt{ 1 + R^2 m^2_A }  }
\Big[(\tilde\gamma\, W'-W \tilde\gamma') - 3  (XV'  -VX') \Big]
+ A_rB_r'-B_rA_r' \Big) \ , \la{513}
\ea
where we put the  terms with $(...)'= \del_\phi (...)$  into the mass  
part anticipating the expansion in Fourier modes in $\phi$.

In general, for two 
mixed 3d scalars  $\Phi_{1},\Phi_{2}$ with canonical kinetic terms and the mass
terms like in \rf{513}
\be
L_{\rm mass} = m_{1}^{2}\Phi_{1}^{2}+m_{2}^{2}\Phi_{2}^{2}+m_{12}^{2} (\Phi_{1}\Phi_{2}'-\Phi_{2}\Phi_{1}')\ ,
\ee
 decomposing the  fields in $\sin(n\phi)$, $\cos(n\phi)$ components 
  the  resulting mass matrix 
\be
{\cal M}^2=\begin{pmatrix}
m_1^2 & 0 & 0 & n m_{12}^{2} \\
 0 & m_1^2 & -n m_{12}^{2} & 0 \\
 0 & -n m_{12}^{2} & m_2^2 & 0 \\
 n m_{12}^{2} & 0 & 0 & m_2^2 
\end{pmatrix}\ , 
\ee
has eigenvalues (cf. \rf{3.17},\rf{3.22})
\be
m^2_\pm  = \tfrac{1}{2} \Big[m_1^2+m_2^2\pm\sqrt{(m_1^2-m_2^2)^{2}+4 n^2 \,
m_{12}^4}\ \Big]\ .
\ee
Assuming  $n\neq 0$, we then obtain the following expansions of 
 masses for the above pairs of mixed bosonic fluctuations:
\ba
\la{5.28}
&m^2_{V,X} =\te  2-\frac{3}{2}kn+\frac{1}{4}k^{2}n^{2}+\big(1-\frac{3}{4}kn\big)\cos^{2}\sigma\ \kk^{2}+\cdots\ ,\\  &
\la{5.30}
m^2_{\tilde \gamma,W} = \te\frac{1}{2}kn+\frac{1}{4}k^{2}n^{2}+\big[\big(1+\frac{1}{4}kn\big)\cos^{2}\sigma-\cos^{4}\sigma\big]\,\kk^{2}
+\cdots\ ,  \\
\la{5.32}
&m^2_{A,B} =\te \frac{1}{2}kn+\frac{1}{4}k^{2}n^{2}+\frac{1}{4}\cos^{2}\sigma\ \kk^{2}+\cdots\ .
\ea
The quadratic  fermionic fluctuations  in \rf{3.222}  are again governed   by the 
 the Dirac operator in  (\ref{4.3}). 
 While in  the $\beta=0$ case in section \ref{sec4}
the mass parameter $m$  was independent of $\sigma$, here 
one finds that in the IIA  string  limit \cite{Aguilera-Damia:2014bqa}
\be
\la{5.33}
m_0 (\sigma) = \tfrac{1}{4}\Big[s_{1}-s_{2}+\frac{\dn\sigma}{\sqrt{1-\kk^{2}}}(s_{3} + 3  s_1 s_2) \Big]\ , 
\ee
where $s_{1},s_{2},s_{3}$ take  $\pm 1$  values giving  masses  of  8  fermionic 2d modes.  
 In the present M2  brane case where the Dirac operator contains also a 
 $\del_\phi$ term 
the  fermion masses are again given by  
(\ref{5.33})  with  an extra 
 universal term $\frac{1}{2}kn $ as in \rf{3.28}, \ie 
 \be  m = m_0(\s)  + \tfrac{1}{2}kn \ , \qquad \qquad n=\pm1, \pm2, \dots\ . \la{329} \ee

\subsection{Expansion of fluctuation frequencies}

Like in section 4 we   can now determine the  leading terms  in the expansion of the corresponding fluctuation frequencies 
$\om_\ell$ in small $\kk$ or, equivalently, in  small $\beta$. 

\subsubsection*{$V,X$ scalars}

The AdS$_2$ conformal dimension $h_{n}$   corresponding to the $\kk=0$  value of the mass 
here  is the  same as in  (\ref{4.19}). Then using \rf{5.28}
the   analogs of  (\ref{4.15})--\rf{4.17} are  
\ba
&-\Big[\partial_{ \tilde\sigma}^{2}-\frac{h_{n}(h_{n}-1)}{\cos^{2}\tilde\sigma}+
\big[\tfrac{1}{2}h_{n}(h_{n}-1)-1+\tfrac{3}{4}kn\big]\, \kk^{2}+\cdots
\Big]\Phi_{h_n,\ell}=\tilde\omega_{\ell}^{2} \Phi_{h_n,\ell} \ ,\la{522}  \\
&\tilde\omega_{\ell}^{2} = (\ell+h_{n})^{2}-\kk^{2}\int_{-\pi/2}^{\pi/2}d\tilde\sigma\, \Phi_{h_{n},\ell}^{2}\big[\tfrac{1}{2}h_{n}(h_{n}-1)
-1+\tfrac{3}{4}kn\big] +\cdots \lp
\ \ \ = (\ell+h_{n})^{2}-\big[\tfrac{1}{2}h_{n}(h_{n}-1)-1+\tfrac{3}{4}kn\big]\, \kk^{2}+\cdots,
\no \\
&
\omega_{\ell} = (\ell+h_{n})(1-\tfrac{1}{4}\kk^{2})-\frac{h_{n}(h_{n}-1)-2+\frac{3}{2}kn}{4(\ell +h_{n})}\, \kk^{2}+\cdots.\la{523}
\ea

\subsubsection*{$\tilde \gamma, W$ scalars}

Here the 
 conformal dimension $h_{n}$ is the  same as in  (\ref{4.12}) and  using  \rf{5.30} 
 we get 
\ba
&-\Big[\partial_{ \tilde\sigma}^{2}-\frac{h_{n}(h_{n}-1)}{\cos^{2}\tilde\sigma}+
\big[\tfrac{1}{2}h_{n}(h_{n}-1)-1-\tfrac{1}{4}kn +\cos^{2}\tilde\sigma+\cdots
  \big]\, \kk^{2}+\cdots
\Big]\Phi_{h_n,\ell}=\tilde\omega_{\ell}^{2} \Phi_{h_n,\ell} \ ,\la{525}  \\
&\tilde\omega_{\ell}^{2} = (\ell+h_{n})^{2}-\kk^{2}\int_{-\pi/2}^{\pi/2}d\tilde\sigma\, \Phi_{h_{n},\ell}^{2}\big[\tfrac{1}{2}h_{n}(h_{n}-1)
-1-\tfrac{1}{4}kn +\cos^{2}\tilde\sigma \big] +\cdots \  \no \\
&\ \ \ \   = (\ell+h_{n})^{2}-\Big[\tfrac{1}{2}h_{n}(h_{n}-1)-1-\tfrac{1}{4}kn + 
\frac{-1+h_{n}(h_{n}-1)+(h_{n}+\ell)^{2}-1}{2(h_{n}+\ell+1)(h_{n}+\ell-1)}
\Big]\, \kk^{2}+\cdots,
\la{5255} \\
&
\omega_{\ell} = (\ell+h_{n})(1-\tfrac{1}{4}\kk^{2})-\Big[\frac{h_{n}(h_{n}-1)-1-\frac{1}{2}kn}{4(\ell + h_{n})}
+\frac{h_{n}(h_{n}-1)}{4(h_{n}+\ell)(h_{n}+\ell+1)(h_{n}+\ell-1)}
\Big]\kk^{2}+\cdots,\no 
\ea
where the last term is the same as in (\ref{4.24}).

\subsubsection*{$A_r,B_r$ scalars}

Again the conformal dimension $h_{n}$  is the  same as in (\ref{4.12})
and using (\ref{5.32}) we get 
\ba
&-\Big[\partial_{ \tilde\sigma}^{2}-\frac{h_{n}(h_{n}-1)}{\cos^{2}\tilde\sigma}+
\big[\tfrac{1}{2}h_{n}(h_{n}-1)-\tfrac{1}{4}\big]\, \kk^{2}+\cdots
\Big]\Phi_{h_n,\ell}=\tilde\omega_{\ell}^{2} \Phi_{h_n,\ell} \ ,\la{5223}  \\
&\tilde\omega_{\ell}^{2} = (\ell+h_{n})^{2}-\kk^{2}\int_{-\pi/2}^{\pi/2}d\tilde\sigma\, \Phi_{h_{n},\ell}^{2}\big[\tfrac{1}{2}h_{n}(h_{n}-1)
-\tfrac{1}{4}\big] +\cdots \lp
\ \ \ = (\ell+h_{n})^{2}-\big[\tfrac{1}{2}h_{n}(h_{n}-1)-\tfrac{1}{4}\big]\, \kk^{2}+\cdots,
\no \\
&
\omega_{\ell} = (\ell+h_{n})(1-\tfrac{1}{4}\kk^{2})-\frac{h_{n}(h_{n}-1)-\frac{1}{2}}{4(\ell +h_{n})}\, \kk^{2}+\cdots.\la{5233}
\ea

\subsubsection*{Fermions}

Using the values of masses in (\ref{5.33}),\rf{329}
 we find the following small $\kk$ expansions of the corresponding mass term in the equation  (\ref{4.29}) for the fermionic  $\om_\ell$ 
\be
\la{5.43}
-\frac{2\KK}{\pi}\frac{\sqrt{1-\kk^{2}}}{\cn(\frac{2\KK}{\pi}\tilde\sigma)} \, m(\tfrac{2\KK}{\pi}\tilde\sigma) = 
\begin{cases}
-\frac{\ha kn}{\cos\tilde\sigma}+\frac{1}{4}(\frac{kn}{2}\pm 1)\cos\tilde\sigma\, \kk^{2}+\cdots\ , \\
-\frac{\ha kn+1}{\cos\tilde\sigma}+\frac{1}{4}(\frac{kn}{2}+1-2)\cos\tilde\sigma\, \kk^{2}+\cdots\ , \qquad \text{(2 \text{modes})} \\
-\frac{\ha kn-1}{\cos\tilde\sigma}+\frac{1}{4}(\frac{kn}{2}-1+2)\cos\tilde\sigma\, \kk^{2}+\cdots\ ,   \qquad \text{(2 \text{modes})} \\
-\frac{\ha kn+1}{\cos\tilde\sigma}+\frac{1}{4}(\frac{kn}{2}+1-1)\cos\tilde\sigma\, \kk^{2}+\cdots\ , \\
-\frac{\ha kn-1}{\cos\tilde\sigma}+\frac{1}{4}(\frac{kn}{2}-1+1)\cos\tilde\sigma\, \kk^{2}+\cdots\ .
\end{cases}
\ee
Comparing with (\ref{4.31}) where we had
the term $-\frac{m}{\cos\tilde\sigma}+\tfrac{1}{4}m\cos\tilde\sigma\ \kk^{2}+\cdots$
 (the $\gamma^{1}$ term from  connection part
  was shown to give no contribution) 
we see that the $\kk^{2}$ terms  in (\ref{5.43})   lead to 
\be
-\frac{m}{\cos\tilde\sigma}+\tfrac{1}{4}(m+\delta m)\cos\tilde\sigma\ \kk^{2}+\cdots,
\ee
with  $\delta m$ shifts   for the $(3+3)+(1+1)=8$ fermions   given by 
\be
\def\arraystretch{1.3}
\begin{array}{cccc}
\toprule
m & \frac{kn}{2}+1 & \frac{kn}{2}-1 & \frac{kn}{2}  \\
\delta m & -2, -2, -1 & 2, 2, 1 &  1,-1 \\
\bottomrule
\end{array}\notag
\ee
Eq. 
 (\ref{4.37}) should then be modified by 
   $\delta m$,   giving 
\ba
\la{5.46}
\omega_{\ell} = (\ell+h_{n})\big(1-\tfrac{1}{4}\kk^{2}\big)+\tfrac{1}{4}\big(\tfrac{1}{2}-h_{n}\mp \delta m\big)\Big[
\frac{h_{n}-\frac{1}{2}}{\ell+h_{n}+\frac{1}{2}}+\frac{2h_{n}-1}{(2\ell+2h_{n}-1)(2\ell+2h_{n}+1)}
\Big]\,\kk^{2}+\cdots\ . 
\ea

\subsection{One-loop   vacuum energy}

Following the discussion in section 4.4  let us 
 start with $n>0$ modes and recall the values of  conformal dimensions 
 in the  zero-cusp AdS$_2$  limit  (labelling fermions as in the $\beta$=0 case)
\be
n>0:\quad\te  h_{\gamma W}=h_{AB} = 1+\frac{kn}{2}\ , \qquad 
h_{VX}=\frac{kn}{2}-1\ , \qquad
h_{\vartheta_\pm} = \frac{kn}{2}\pm 1+\frac{1}{2}\ ,\qquad
h_{\vartheta'} = \frac{kn}{2}+\frac{1}{2}\ .
\ee
We then  doing the sum over $\ell$ in \rf{446} we find
\ba
E_n^{\gamma W} =& 2\times\Big[\tfrac{1}{2}\zeta(-1,h_{\gamma W})(1-\tfrac{1}{4}\kk^{2})-\tfrac{1}{8}\big[
h_{\gamma W}(h_{\gamma W}-1)-1-\tfrac{kn}{2}\big]\zeta(1,h_{\gamma W})\kk^{2}-\tfrac{1}{16}\kk^{2}+\cdots\Big]\ , \no\\
E_n^{AB} = &4\times\Big[\frac{1}{2}\zeta(-1,h_{AB})(1-\tfrac{1}{4}\kk^{2})-\tfrac{1}{8}\big[
h_{AB}(h_{AB}-1)-\tfrac{1}{2}\big]\zeta(1,h_{AB})\kk^{2}+\cdots\Big], \la{535}\\
E_n^{VX} =&2\times\Big[\tfrac{1}{2}\zeta(-1,h_{VX})(1-\tfrac{1}{4}\kk^{2})-\tfrac{1}{8}\big[
h_{VX}(h_{VX}-1)-2+\tfrac{3}{2}kn\big]\zeta(1,h_{VX})\kk^{2}+\cdots\Big]\ ,\no \\
E_n^{\vartheta_{\pm}} =&2\times\Big[
-\tfrac{1}{2}\zeta(-1,h_{\vartheta_{\pm}})(1-\tfrac{1}{4}\kk^{2})+\tfrac{1}{8}(\tfrac{1}{2}-h_{\vartheta_{\pm}}\pm 2)
\big[(\tfrac{1}{2}-h_{\vartheta_{\pm}})
\zeta(1,h_{\vartheta_{\pm}}+\tfrac{1}{2})-\tfrac{1}{2}\big]\kk^{2}+\cdots
\Big]\lp
+1\times\Big[
-\tfrac{1}{2}\zeta(-1,h_{\vartheta_{\pm}})(1-\tfrac{1}{4}\kk^{2})+\tfrac{1}{8}(\tfrac{1}{2}-h_{\vartheta_{\pm}}\pm 1)
\big[(\tfrac{1}{2}-h_{\vartheta_{\pm}})
\zeta(1,h_{\vartheta_{\pm}}+\tfrac{1}{2})-\tfrac{1}{2}\big]\kk^{2}+\cdots
\Big]\ , \no\qquad  \\ 
E_n^{\vartheta'} =&1 \times\Big[
-\tfrac{1}{2}\zeta(-1,h_{\vartheta'})(1-\tfrac{1}{4}\kk^{2})+\tfrac{1}{8}(\tfrac{1}{2}-h_{\vartheta'}+1)
\big[(\tfrac{1}{2}-h_{\vartheta'})
\zeta(1,h_{\vartheta'}+\tfrac{1}{2})-\tfrac{1}{2}\big]\kk^{2}+\cdots
\Big]\lp
+1\times\Big[
-\tfrac{1}{2}\zeta(-1,h_{\vartheta'})(1-\tfrac{1}{4}\kk^{2})+\tfrac{1}{8}(\tfrac{1}{2}-h_{\vartheta'}-1)
\big[(\tfrac{1}{2}-h_{\vartheta'})
\zeta(1,h_{\vartheta'}+\tfrac{1}{2})-\tfrac{1}{2}\big]\kk^{2}+\cdots
\Big]\ .\no
\ea
Similarly, for $n <0$ we get 
\ba
n<0:\quad&\te  h_{\gamma W}=h_{AB} = -\frac{kn}{2}\ , \qquad 
h_{VX}=2-\frac{kn}{2}\ , \quad
h_{\vartheta_\pm} = -\frac{kn}{2}\mp 1+\frac{1}{2}\ ,\quad
h_{\vartheta'} = -\frac{kn}{2}+\frac{1}{2}\ ,\la{537}
\\
E_n^{\gamma W} =&2\times\Big[\tfrac{1}{2}\zeta(-1,h_{\gamma W})(1-\tfrac{1}{4}\kk^{2})-\tfrac{1}{8}\big[
h_{\gamma W}(h_{\gamma W}-1)-1-\tfrac{kn}{2}\big]\zeta(1,h_{\gamma W})\kk^{2}-\tfrac{1}{16}\kk^{2}+\cdots\Big]\ , \no \\
E_n^{AB} =&4\times\Big[\frac{1}{2}\zeta(-1,h_{AB})(1-\tfrac{1}{4}\kk^{2})-\tfrac{1}{8}\big[
h_{AB}(h_{AB}-1)-\tfrac{1}{2}\big]\zeta(1,h_{AB})\kk^{2}+\cdots\Big]\ , \la{538} \\
E_n^{VX} =&2\times\Big[\tfrac{1}{2}\zeta(-1,h_{VX})(1-\tfrac{1}{4}\kk^{2})-\tfrac{1}{8}\big[
h_{VX}(h_{VX}-1)-2+\tfrac{3}{2}kn\big]\zeta(1,h_{VX})\kk^{2}+\cdots\Big]\ , \no \\
E_n^{\vartheta_{\pm}} =&2\times\Big[
-\tfrac{1}{2}\zeta(-1,h_{\vartheta_{\pm}})(1-\tfrac{1}{4}\kk^{2})+\tfrac{1}{8}(\tfrac{1}{2}-h_{\vartheta_{\pm}}\mp 2)
\big[(\tfrac{1}{2}-h_{\vartheta_{\pm}})
\zeta(1,h_{\vartheta_{\pm}}+\tfrac{1}{2})-\tfrac{1}{2}\big]\kk^{2}+\cdots
\Big]\no \\ &
+1\times\Big[
-\tfrac{1}{2}\zeta(-1,h_{\vartheta_{\pm}})(1-\tfrac{1}{4}\kk^{2})+\tfrac{1}{8}(\tfrac{1}{2}-h_{\vartheta_{\pm}}\mp 1)
\big[(\tfrac{1}{2}-h_{\vartheta_{\pm}})
\zeta(1,h_{\vartheta_{\pm}}+\tfrac{1}{2})-\tfrac{1}{2}\big]\kk^{2}+\cdots
\Big]\ ,  \no \qquad\qquad \\ 
E_n^{\vartheta'} =&1\times\Big[
-\tfrac{1}{2}\zeta(-1,h_{\vartheta'})(1-\tfrac{1}{4}\kk^{2})+\tfrac{1}{8}(\tfrac{1}{2}-h_{\vartheta'}+1)
\big[(\tfrac{1}{2}-h_{\vartheta'})
\zeta(1,h_{\vartheta'}+\tfrac{1}{2})-\tfrac{1}{2}\big]\kk^{2}+\cdots
\Big]\no \lp
+1\times\Big[
-\tfrac{1}{2}\zeta(-1,h_{\vartheta'})(1-\tfrac{1}{4}\kk^{2})+\tfrac{1}{8}(\tfrac{1}{2}-h_{\vartheta'}-1)
\big[(\tfrac{1}{2}-h_{\vartheta'})
\zeta(1,h_{\vartheta'}+\tfrac{1}{2})-\tfrac{1}{2}\big]\kk^{2}+\cdots
\Big]\ .\no
\ea
The difference between the expressions for $n>0$ and $n<0$ is only in the sign
 of  $\delta m$ in (\ref{5.46})
which is due to the fact that we should use $(\psi^{1}, \psi^{2})$ spinors in the $n<0$ case, \cf  (\ref{4.28}).

Note that  for 
 $n\neq 0$ we do not  have massless fermions and thus  there is no ambiguity in the  choice of their quantization. This is 
to be compared with the  string theory, \ie $n=0$,  case considered  in 
\cite{Aguilera-Damia:2014bqa} where  for $\alpha=0, \ \beta\neq0 $ there were $\kk$ corrections to  massless fermions
and one had to choose a quantization consistent with $\N=6$ supersymmetry
in the zero cusp limit. This subtlety is not present for
  $n\neq 0$.

Adding all mode contributions together   for $n=0$ we  find
 $E_0=  \frac{1}{4}\kk^{2} + \mc O(\kk^2) $ as in
\cite{Aguilera-Damia:2014bqa} (cf. \rf{455}).
 The  total sum over $n$ 
is finite and is  given by  the   same expression 
  as in the $\beta=0$ case in  \rf{456}
\be\la{560}
E= \sum^\infty_{n=-\infty}E_{n} = \frac{\pi}{2k}\cot\frac{2\pi}{k}\ \kk^{2}+\mc O(\kk^4) \ . 
\ee
Taking into account \rf{a8}, \ie  that in the small  $\beta$ cusp   limit  $\kk^{2} = -\frac{\beta^{2}}{\pi^{2}} + ...$ 
(while in $\beta=0$ case we had $\kk^{2} = \frac{\alpha^{2}}{\pi^{2}}$, 
(\cf  \rf{a7})   we confirm that the $\alpha^2$ and $-\beta^2$ 
terms  have the coefficient \rf{1.22},  in agreement with \rf{1.5}. 
The same   conclusion is reached also in the special $k=1,2$ cases.

\iffa 
and comparing to 
we obtain the same expression as in $\beta=0$ case,
\be
B_{\abjm}^{\rm one-loop} = -\frac{1}{2\pi k}\cot\frac{2\pi}{k},
\ee
reproducing (\ref{1.10}), and confirming the BPS structure (\ref{1.5}), at least in the two extreme cases $\alpha=0$ or $\beta=0$.
\fi 

\section*{Acknowledgements}

AAT  is grateful to S. Giombi for  many discussions and   
collaboration on \cite{Giombi:2024itd}   where it was conjectured  that the 
non-planar cotangent term in the  \brem function \rf{1.10} 
 can be derived  by  quantizing the M2  brane.  We also thank V. Forini, V. M. Puletti, and O. Ohlsson Sax for clarifications related to  their work \cite{Forini:2012bb}.
 MB is supported by the INFN grant GAST.
AAT is  supported by the STFC grant ST/T000791/1.



\bibliography{BT-Biblio}
\bibliographystyle{JHEP-v2.9}
\end{document}